\documentclass{article}

\usepackage{graphics} 
\usepackage{epsfig} 
\usepackage{mathptmx} 
\usepackage{times} 
\usepackage{amsmath} 
\usepackage{amssymb}  
\usepackage{verbatim}
\usepackage{color}

\date{}

\begin{document}


\title{\LARGE {\bf
Collision Course for Objects Moving on a Spherical Manifold} \\
}

\author{Animesh Chakravarthy* and Debasish Ghose** \\
*University of Texas at Arlington, TX 76019, USA\\
**Indian Institute of Science, Bangalore 560 012, India}



\maketitle


\begin{abstract}
In this paper, we address the problem of predicting collision for objects moving on the surface of a spherical manifold.  Toward this end, we develop the notion of a collision triangle on such manifolds.  We use this to determine analytical conditions governing the speed ratios and direcions of motion of objects that lead to collisions on the sphere.  We first develop these conditions for point objects and subsequently extend this to the case of circular patch shaped objects moving on the sphere.

\end{abstract}



\section{INTRODUCTION}

Predicting collision between objects moving in space has been an area of research for the past few years and has given rise to a rich literature. One of the fundamental works in this direction addresses the identification of initial velocity directions that lead to collision between objects moving in straight-lines. This leads to the idea of collision cones which is the set of velocity directions of an object that on a plane that leads to collision with another object. These objects are not necessarily points or circles but can also have arbitrary irregular or non-convex shapes \cite{smc1}. Other researchers have proposed different ways of looking at collision dynamics between objects moving on a plane too \cite{vo1}. However, there is no work in the literature that address the prediction of collision between objects moving on spherical manifolds. 

There is some literature on different applications of objects moving on spherical surfaces but they do not discuss the problem of collision prediction.  In \cite{okoloko}, the authors pose the problem of consensus on a spherical manifold using LMIs and consider collision avoidance by creating separation among point objects via semi-definite programming.  In \cite{darbha}, the authors propose motion planning algorithms for Dubins’ vehicles on spherical manifolds. Much of the basic mathematics required to study motion planning on spheres is based on spherical trigonometry \cite{brum}.

In this paper, we develop a self-contained theory of predicting collisions among objects moving on a sphere and extend it to objects of circular shapes. We define collision cones on spherical surfaces and show that these are generalizations of collision cones on planes. We also derive many new results that have no analogous results in the planar case.

\section{COLLISION COURSE IN EUCLIDEAN SPACE}

In this section, we briefly review the basic collision course conditions for objects moving with constant velocities in $R^2$ and $R^3$, respectively.    
Consider two point objects $A$ and $B$ moving on a plane.  Assume that $A$ and $B$ are both moving with constant velocities.  Let $V_A$ and $V_B$ represent the speeds of $A$ and $B$, and let $\alpha$ and $\beta$ represent their respective heading angles.  By virtue of the constant velocity assumption, the quantities $V_A,V_B,\alpha,\beta$ are all constants.  Let $r(t)$ represent the distance between $A$ and $B$ and $\theta(t)$ represent the bearing angle of $B$ with respect to $A$.

\begin{figure}
\centering
\includegraphics[width=2.5in]{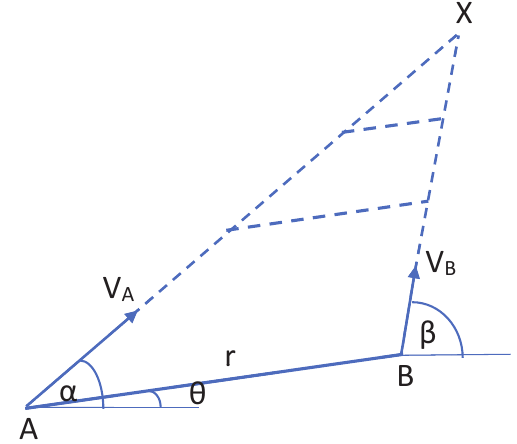}
\includegraphics[width=2.5in]{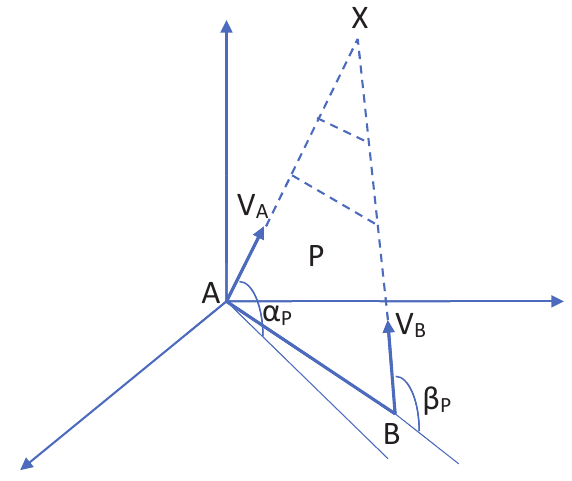}
\caption{Collision Triangle on (a) an Euclidean plane, (b) in Euclidean space}
\label{fig_coll2D}
\end{figure}

At any time $t$, the objects $A$ and $B$ are said to be on a collision course, if their current trajectories are such that there exists a future time $t_f$ at which they will collide.  This can be conveniently visualized using the notion of a collision triangle.  Refer Fig \ref{fig_coll2D}, which shows the engagement geometry between $A$ and $B$, as well as the ensuing collision triangle $ABX$ that is formed when they are on a collision course.  

When $A$ and $B$ are on trajectories that lie on a collision triangle, then two properties are satisfied: (a) The lines joining the instantaneous positions of $A$ and $B$ at successive times are all parallel to each other, and (b) The distance between $A$ and $B$ is continuously decreasing.  This is schematically demonstrated in the collision triangle of Fig \ref{fig_coll2D}(a), where the lines $AB$, $A'B'$, $A''B''$ are all parallel to each other, and have progressively decreasing lengths.  

Let $\vec{V}_R = \vec{V}_B - \vec{V}_A$ represent the relative velocity of $B$ with respect to $A$.  Now, define two quantities $V_\theta$ and $V_r$, as:
\begin{eqnarray}
V_\theta &=& V_B \sin(\beta-\theta) - V_A \sin(\alpha-\theta) \label{Vtheta_eqn} \\
V_r &=& V_B \cos(\beta-\theta) - V_A \cos(\alpha-\theta) \label{Vr_eqn}
\end{eqnarray}    
It is evident that at time $t$, $V_\theta(t)$ represents the component of the relative velocity $\vec{V}_R$ acting normal to the line joining the positions of $A$ and $B$, and $V_r(t)$ represents the component of the relative velocity that acts along this line.  Thus, the condition $V_\theta(t)=0$ implies that the line $AB$ is non-rotating, and the condition $V_r(t) < 0$ implies that the length of the line $AB$ is reducing.  We can now state the following Lemma.

{\bf Lemma 1:} Consider two point objects $A$ and $B$, both moving with constant velocities on a plane.  Then, $A$ and $B$ are on a collision course if and only if the conditions $V_\theta=0$ and $V_r < 0$ are satisfied.  

When $A$ and $B$ are moving with constant velocities in $R^3$, we can still use Lemma 1 to determine the collision conditions as follows.  Shift the velocity vector $\vec{V}_B$ to $A$ and then construct a plane $P$ that contains the velocity vectors $\vec{V}_A$ and $\vec{V}_B$.  Let $\hat{V}_A$ and $\hat{V}_B$ represent the unit velocity vectors of $A$ and $B$, respectively, and $\hat{r}(t)$ represent the unit vector along the LOS.  Refer Fig \ref{fig_coll2D}(b).  If $A$ and $B$ are on a collision course, then the plane $P$ will contain a collision triangle.  Defining $\alpha_P$ and $\beta_P$  as the heading angles of $A$ and $B$ measured on the plane $P$, and $\theta_P$ as the bearing angle measured on this plane, we 
can see that $\alpha_P - \theta_P = \cos^{-1}(\hat{V}_A.\hat{r})$, $\beta_P-\theta_P = \cos^{-1}(\hat{V}_B.\hat{r})$.  

We can then use (\ref{Vtheta_eqn}) and (\ref{Vr_eqn}) to determine $V_{\theta P}$ and $V_{rP}$ on $P$.  Then, $A$ and $B$ will be on a collision course if and only if the conditions $V_{\theta P}=0$ and $V_{rP}<0$ are satisfied.   
The condition representing non-rotation of the LOS can be mathematically written as the following conditions being satisified for all time $t$: 
\begin{eqnarray}
\hat{r}. (\hat{V}_A \times \hat{V}_B) = 0  \label{coll_cond1}\\
V_B \sin \underbrace{(\cos^{-1}(\hat{V}_B.\hat{r}))}_{\beta_P-\theta_P} - V_A \sin \underbrace{(\cos^{-1}(\hat{V}_A.\hat{r}))}_{\alpha_P-\theta_P} = 0 \label{coll_cond2}
\end{eqnarray}
where, the first equation ensures that the LOS remains on a constant plane $P$, and the second equation ensures that successive LOS on $P$ are all parallel to one another. 

An important point to note is that Lemma 1, in conjunction with (\ref{Vtheta_eqn}), (\ref{Vr_eqn}) enables the prediction of collision without knowledge of the collision point $X$, and the time to collision.  Our objective in this paper is to determine corresponding collision conditions for objects moving on the surface of a sphere.

\section{PRELIMINARIES}

We briefly provide some preliminary information on spherical geometry.  A great circle on a sphere is a circle drawn on the surface of a sphere, such that the center of the circle is also the center of the sphere.  The shortest distance between two points on the surface of a sphere is equal to the smaller of the two arc lengths of the great circle passing through those two points.  A spherical triangle is a triangle formed by three intersecting great circles.  A spherical lune is the portion of a sphere bounded by two half great circles.  

Unlike a planar triangle, the sum of the angles of a spherical triangle is not a constant.  The sum of angles of a spherical triangle lies between 180 degrees and 540 degrees.  Let $x,y,z$ represent the arc lengths of the sides of the spherical triangle and $X,Y,Z$ represent the corresponding angles of the triangle.  A spherical triangle satisfies the following trigonometric identities:
\begin{eqnarray}
\frac{\sin X}{\sin x} &=& \frac{\sin Y}{\sin y} ~~=~~ \frac{\sin Z}{\sin z}  \label{sine_rule_basic}\\
\cos{z} &=& \cos{x} \cos{y} + \sin{x} \sin{y} \cos{Z} \label{cosine_rule1_basic}\\
\cos{Z} &=& -\cos{X} \cos{Y} + \sin{X} \sin{Y} \cos{z} \label{cosine_rule2_basic}
\end{eqnarray} 
In the above, (\ref{sine_rule_basic}) is the spherical sine rule, (\ref{cosine_rule1_basic}) is the first spherical law of cosines, and (\ref{cosine_rule2_basic}) is the second spherical law of cosines.  We note that alternative versions of (\ref{cosine_rule1_basic}) and (\ref{cosine_rule2_basic}) can be written by performing a cyclic permutation of the quantities in each of these equations. 

\section{ENGAGEMENT GEOMETRY ON A SPHERE}

\begin{figure}
\centering
\includegraphics[width=2.5in]{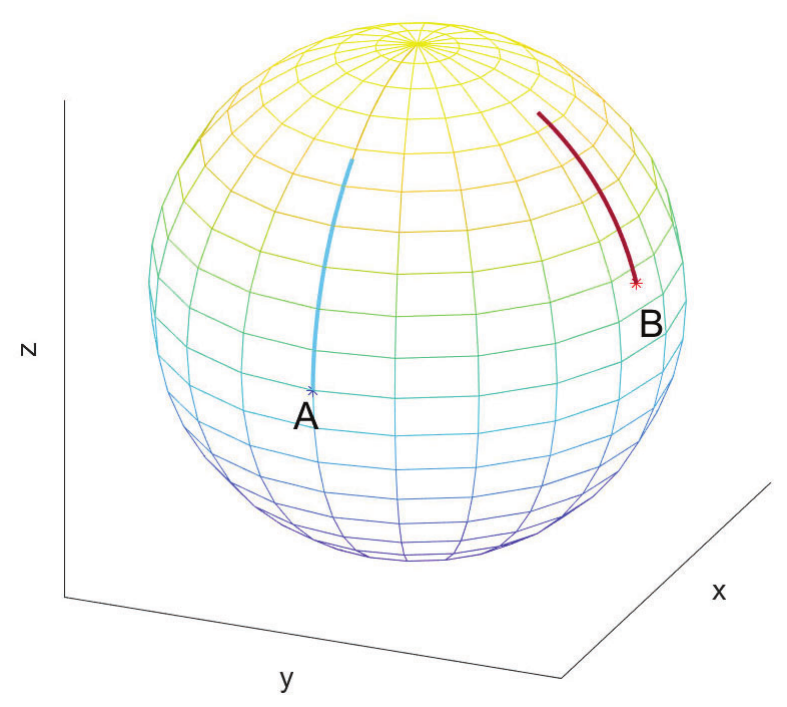}
\includegraphics[width=2.5in]{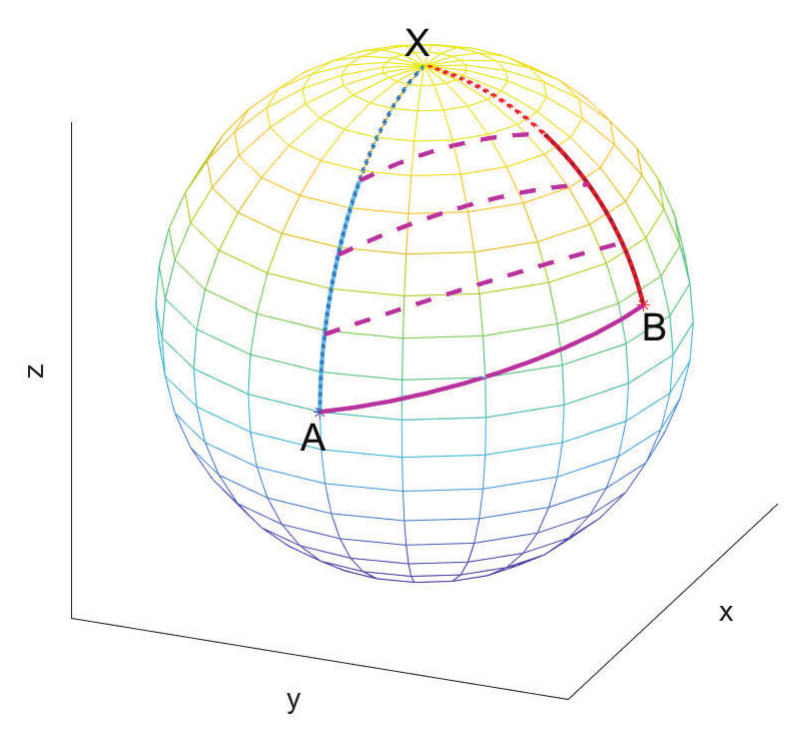}
\caption{(a) $A$ and $B$ moving on the surface of a sphere, (b) Collision Triangle on the surface of a sphere}
\label{fig_sphere}
\end{figure}

We now consider the point objects $A$ and $B$ moving on the surface of a sphere of radius $R$.  We note that in Euclidean space, the shortest distance between two points is a straight line and thererfore, the assumption that $A$ and $B$ move along straight lines is reasonable in the sense that $A$ and $B$ are both taking the shortest paths to their respective destinations.  In a similar vein, since the shortest distance between two points on a sphere lies on the great circle that passes through those points, we therefore consider the scenario where $A$ and $B$ are moving along two distinct great circles, as shown in Fig \ref{fig_sphere}(a).

Let $A$ and $B$ move along two distinct great circles $C_A$ and $C_B$, respectively, on the surface of a sphere.  The points of intersection of the two great circles are referred to as poles.  Without loss of generality, we will define the first pole that $A$ encounters, as the North Pole.  The other pole will be referred to a the South Pole.  If $B$’s initial position is such that the first pole that it encounters is the North Pole, then both $A$ and $B$ are said to be in the same Lune (referred to as Lune $1$), which is the lune that leads to the North Pole.  See Fig ().    
If $B$’s initial position is such that the first pole that it encounters is the South Pole, then $A$ and $B$ are in different lunes (referred to as Lune $1$ and Lune $2$), and where Lune $1$ is the lune that leads to the North Pole.  We also demarcate each lune into two half-lunes (as schematically shown in Fig \ref{fig_spherical5a}), and refer to these four half-lunes as $Q_I$, $Q_{II}$, $Q_{III}$ and $Q_{IV}$.  Note that the arc length of each of the two lunes is $\pi$ radians and that of each of $Q_{I},\cdots,Q_{IV}$ is $\pi/2$ radians.
Let $V_A$ and $V_B$ represent the speeds of $A$ and $B$, respectively, and assume that both $V_A$ and $V_B$ are constant in time.

\section{COLLISION COURSE ON A SPHERE IN THE VELOCITY SPACE}

Refer Fig \ref{fig_sphere}(a), which shows $A$ and $B$ moving on the surface of a sphere.  If $A$ and $B$ are on a collision course, then their individual trajectories, projected into future time, will intersect at the point of collision $X$.  This is schematically shown in Fig \ref{fig_sphere}(b). 
Also shown in this figure is the spherical LOS at time $t=0$, as well as at a few subsequent times till collision.  Since $A$ and $B$ are both moving along great circles, and since the shortest distance between $A$ and $B$ also lies on a great circle, therefore the geometric entity $ABX$ is a spherical triangle.  

\subsection{Sides of the spherical triangle}

Such a spherical triangle is depicted in Fig \ref{fig_spherical2}.  
The vertices of the triangle are given by $A$, $B$ and $X$, where $X$ is the collision point.  Assume that the engagement starts at time $t=0$ and the collision occurs at some time $t_f$.  Then at time $t=0$, the sides of the spherical triangle have arc lengths $\displaystyle\frac{V_A t_f}{R}$, $\displaystyle\frac{V_B t_f}{R}$, $\displaystyle\frac{s(0)}{R}$.  At any intermediate time $t$, the three sides of the spherical triangle $AX$, $BX$ and $AB$ have arc lengths, $\displaystyle\frac{V_A(t_f-t)}{R}$, $\displaystyle\frac{V_B(t_f-t)}{R}$, and $\displaystyle\frac{s(t)}{R}$, respectively, as in Fig Fig \ref{fig_spherical2}.  

\subsection{Angles of the spherical triangle}

\begin{figure}[h]
\centering
\includegraphics[width=3in]{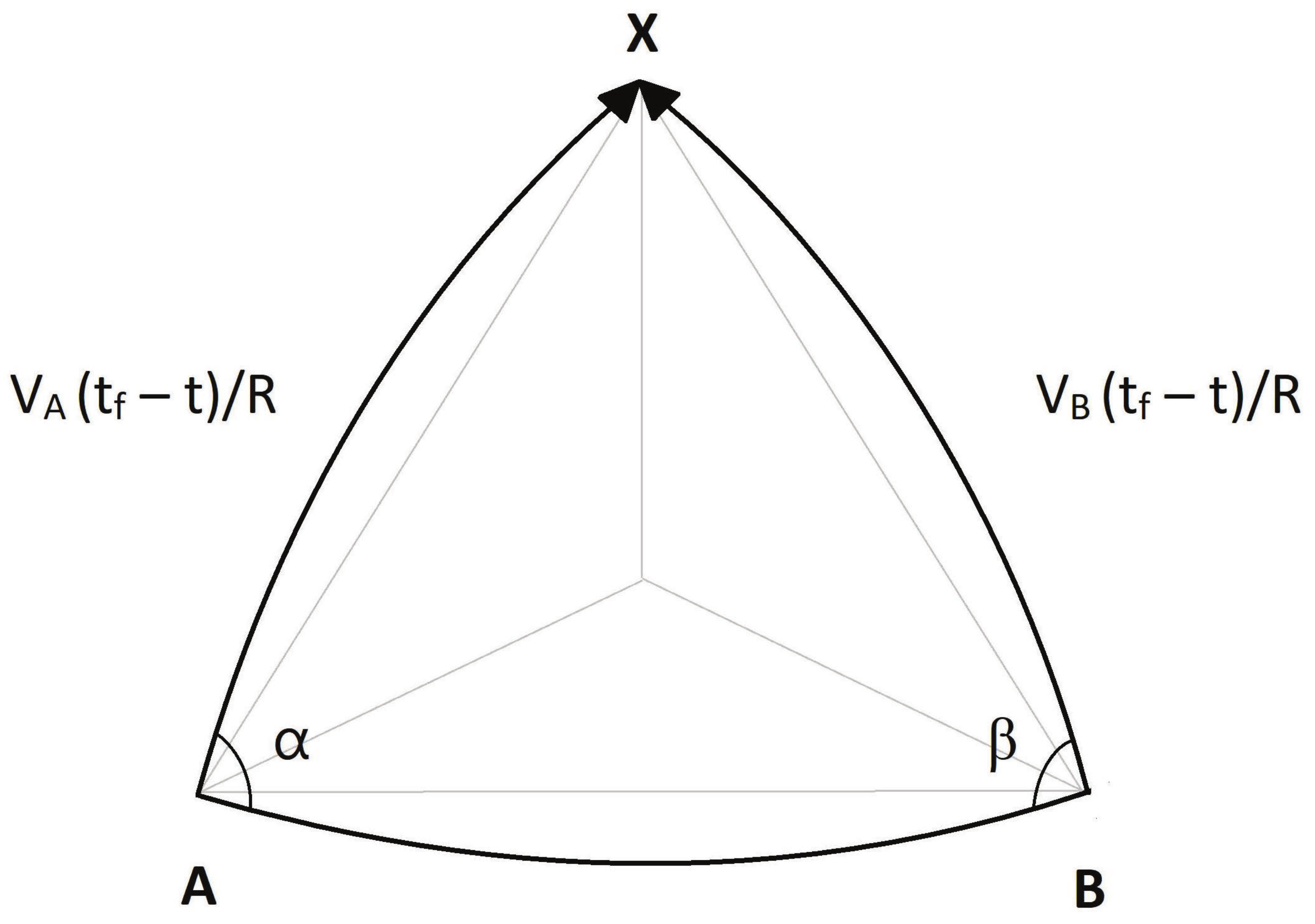}
\caption{Spherical Triangle Collision Geometry}
\label{fig_spherical2}
\end{figure}

We next look at determining the three angles of the spherical triangle of Fig \ref{fig_spherical2}.  
The angle $\alpha(t)$ can be computed from the dot product of the tangents to arcs $AX$ and $AB$, computed at $A$.  
Along similar lines, the angle $\beta(t)$ is found from the dot product of the tangents to arcs $BX$ and $AB$, computed at $B$.

After computing the angles $\alpha$ and $\beta$, we now turn our attention on the third angle $\gamma$, which is the angle between $C_A$ and $C_B$.  If this was a planar triangle, $\gamma$ would be simply given by $180-(\alpha+\beta)$.  For a spherical triangle however, this is not the case.  At time $t$, let the spherical distance be $s(t)$, $\alpha(t)$ represent the angle made by $C_A$ with the spherical LOS and $\beta(t)$ represent the angle made by $C_B$ with the spherical LOS.  Then, the angle $\gamma$ can be computed using the second spherical law of cosines (\ref{cosine_rule2_basic}), to be the following:
\begin{equation}
\cos{\gamma} =  \sin(\alpha(t)) \sin(\beta(t)) \cos(s(t)/R) -\cos(\alpha(t)) \cos(\beta(t))  \label{cosine2n}
\end{equation}   
Note that the above equation enables the computation of $\gamma$ using the instantaneous information of $\alpha(t),\beta(t),s(t)$.  Furthermore, since the left hand side of the above equation is a constant, while the right hand side contains terms that are individually functions of time, we can infer the following:

\noindent {\bf Lemma 2}: The following expression holds for all $t$ till either $A$ and/or $B$ reach $X$.   
\begin{eqnarray}
\sin(\alpha(t)) \sin(\beta(t)) \cos(s(t)/R) -\cos(\alpha(t)) \cos(\beta(t)) = \rm{constant}  \label{cosine}
\end{eqnarray}

\noindent {\bf Proof}: Follows from the above.  $\blacksquare$ 

\section{COLLISION CONDITIONS FOR POINT OBJECTS MOVING ON A SPHERICAL MANIFOLD}

\begin{figure}[h]
\centering
\includegraphics[width=2.5in]{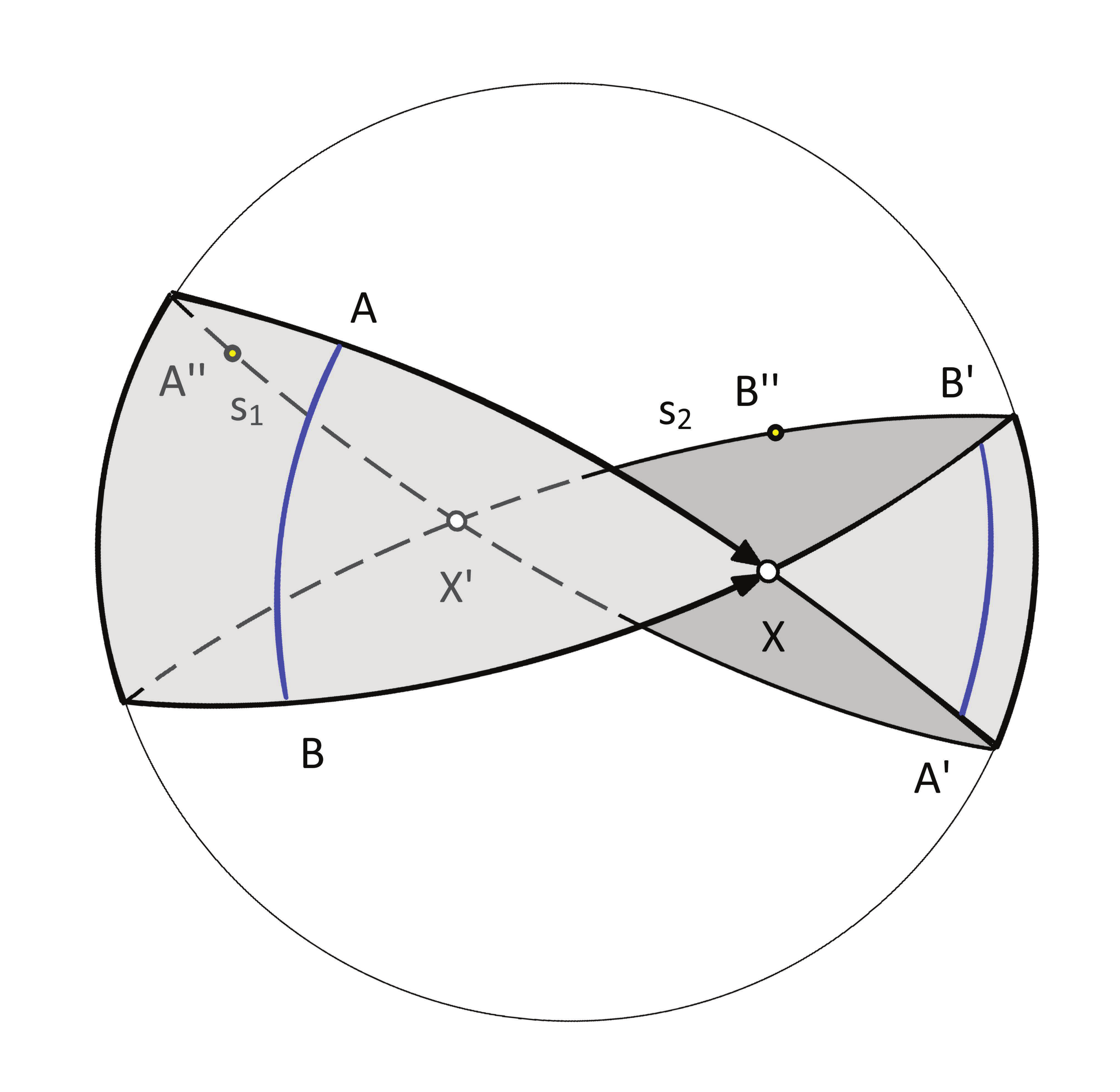}
\includegraphics[width=2.5in]{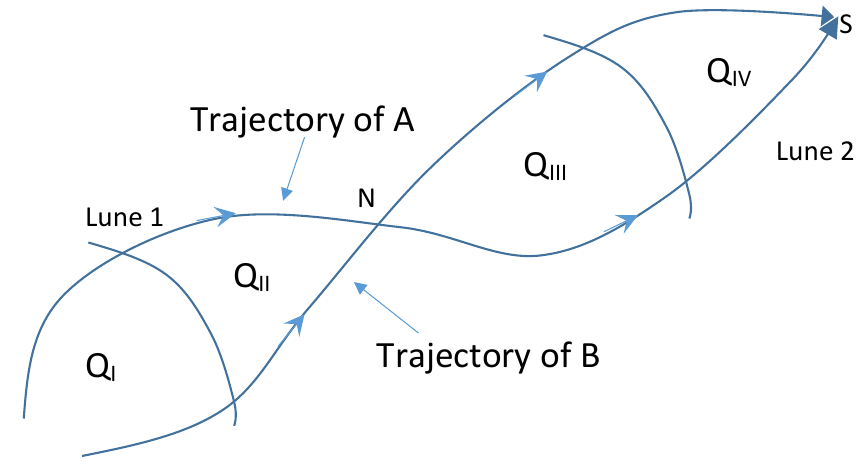}
\caption{(a) Illustration of two lunes, (b) Definition of $Q_I$, $Q_{II}$, $Q_{III}$, $Q_{IV}$}
\label{fig_spherical5a}
\end{figure}

Now that we have defined the sides and angles of the spherical collision triangle, we will use them to determine the collision conditions on a sphere.  Refer Fig \ref{fig_spherical5a}(a), which shows two lunes, referred to as Lune 1 and Lune 2, respectively, formed by the two great circles $C_A$ and $C_B$.  The intersection points of these two lunes are denoted by $N$ and $S$.  Fig \ref{fig_spherical5a}(b)   
provides an alternative view of Fig \ref{fig_spherical5a}(a), in that it ``unfolds'' the two lunes off the surface of the sphere and lays them on a plane.  Note that Fig \ref{fig_spherical5a}(b) also demarcates the two lunes into regions referred to as $Q_I$, $Q_{II}$, $Q_{III}$ and $Q_{IV}$.  

Let $\nu = \frac{V_A}{V_B}$ represent the ratio of speeds of $A$ and $B$, and let $\alpha_0$, $\beta_0$ and $s_0$ represent the initial values of $\alpha$, $\beta$ and $s$, respectively.  We define the quantities $\bar{Z}_{\alpha0}$ and $\bar{Z}_{\beta0}$ as follows:
\begin{eqnarray}
\bar{Z}_{\alpha0} &=& \sin^{-1}\Big[\frac{\sin \alpha_0 \sin (s_0/R)}{\sqrt{1 - [\sin \alpha_0 \sin\beta_0 \cos s_0 - \cos \alpha_0 \cos\beta_0]^2}} \Big]   \label{zalpha} \\
\bar{Z}_{\beta0} &=& \sin^{-1} \Big[\frac{\sin \beta_0 \sin (s_0/R)}{\sqrt{1 - [\sin \alpha_0 \sin\beta_0 \cos s_0 - \cos \alpha_0 \cos\beta_0]^2}}\Big] \label{zbeta} 
\end{eqnarray}
Additionally, we define quantities $Z_{\alpha0}$ and $Z_{\beta0}$ follows.  At $t=t_0$:
\begin{eqnarray}
&&\rm{If}~ A \in Q_{I}, ~\rm{then}~ Z_{\beta0} = \pi - \bar{Z}_{\beta0} \label{zb1}\\
&&\rm{If}~ A \in Q_{II}, ~\rm{then}~ Z_{\beta0} = \bar{Z}_{\beta0} \label{zb2} \\
&&\rm{If}~ B \in Q_{I}, ~\rm{then}~ Z_{\alpha0} = \pi - \bar{Z}_{\alpha0} \label{za1}\\
&&\rm{If}~ B \in Q_{II}, ~\rm{then}~ Z_{\alpha0} = \bar{Z}_{\alpha0} \label{za2} 
\end{eqnarray}
 We can now state the following:

\noindent {\bf Theorem 1}: Consider two point objects $A$ and $B$ moving with constant speeds along two great circles on a sphere.  Assume they are initially in the same lune.  Then 
$A$ and $B$ will collide if and only if their speed ratio satisfies the following equation:
\begin{equation}
\nu = \frac{Z_{\beta0}+q\pi}{Z_{\alpha0}+ p \pi}, ~~p,q = 0,1,2,\cdots  \label{speed_ratio2}
\end{equation}
where, $p,q$ are integers whose values depend on the initial directions of travel of $A$ and $B$, as well as the eventual pole of collision, as follows:
(a)	If $A$ and $B$ are both initially moving northwards, then for them to collide at $N$, $p,q$ both need to be even.  
(b)	If $A$ and $B$ are both initially moving northwards, then for them to collide at $S$, $p,q$ both need to be odd. 
(c)	If $A$ is initially moving northwards and $B$ initially moving southwards, then for them to collide at $N$, $q$ needs to be even, $p$ needs to be odd. 
(d)	If $A$ is initially moving northwards and $B$ initially moving southwards, then for them to collide at $S$, $q$ needs to be odd, $p$ needs to be even. 

{\bf Proof:} Let $X$ represent the point of collision, where $X$ is one of the two poles (either $N$ or $S$).  Then, $A$ and $B$ will collide at $X$ if and only if the ratio of their intial distances to $X$ is equal to their speed ratio.  If $Z_{A0}$ and $Z_{B0}$ represent the initial distances of $A$ and $B$, respectively, to $X$, then the speed ratio for collision to occur is $\nu = \frac{Z_{A0}}{Z_{B0}}$.

Using the spherical sine rule, we have:
\begin{equation}
\frac{\sin (Z_{A0}/R)}{\sin \beta_0} = \frac{\sin (Z_{B0}/R)}{\sin \alpha_0} = \frac{\sin (s_0/R)}{\sin \gamma} \label{sine_rule}
\end{equation}
Collecting the first and third terms in (\ref{sine_rule}), we get:
\begin{eqnarray}
\frac{\sin(Z_{A0}/R)}{\sin \beta_0} = \frac{\sin(s_0/R)}{\sin \gamma} 
\Rightarrow 
Z_{A0}/R = \underbrace{\sin^{-1} \Big[\frac{\sin \beta_0 \sin (s_0/R)}{\sin \gamma} \Big]}_{\bar{Z}_{\beta0}} + q\pi, q = 0,1, \cdots, \label{VA_eqn}
\end{eqnarray}
In the above, $q$ is an index, that is introduced to cater to the fact that in the computation of $Z_{\beta0}$, only the principal value of $\sin^{-1}(.)$ is taken, and therefore $Z_{\beta0} \in [-\frac{\pi}{2},\frac{\pi}{2}]$.  Thus, the case of $q=0$ represents the situation when $A$ collides with $B$ at $X$ at the very first instant that $A$ reaches $X$.  In other cases, $q$ represents the number of half-cycles completed by $A$ before it collides with $B$ at $X$.    
Similarly, considering the second and third terms in (\ref{sine_rule}) and performing a similar re-arrangement of the terms, we get:
\begin{equation}
 Z_{B0}/R = \underbrace{\sin^{-1} \Big[\frac{\sin \alpha_0 \sin (s_0/R)}{\sin \gamma} \Big]}_{\bar{Z}_{\alpha0}} + p\pi, p = 0,1, \cdots, \label{VB_eqn}
\end{equation}  
where $p$ represens the number of half-cycles completed by $B$ before it collides with $A$ at $X$.
In the definitions of both $\bar{Z}_{\alpha0}$ and $\bar{Z}_{\beta0}$, we then use  (\ref{cosine2n}) to substitute $\sin{\gamma}$ in terms of other quantities.  Using these in  (\ref{VA_eqn}) and (\ref{VB_eqn}), we get (\ref{speed_ratio2}).   $\blacksquare$

\begin{figure}
\centering
\includegraphics[width=2.5in]{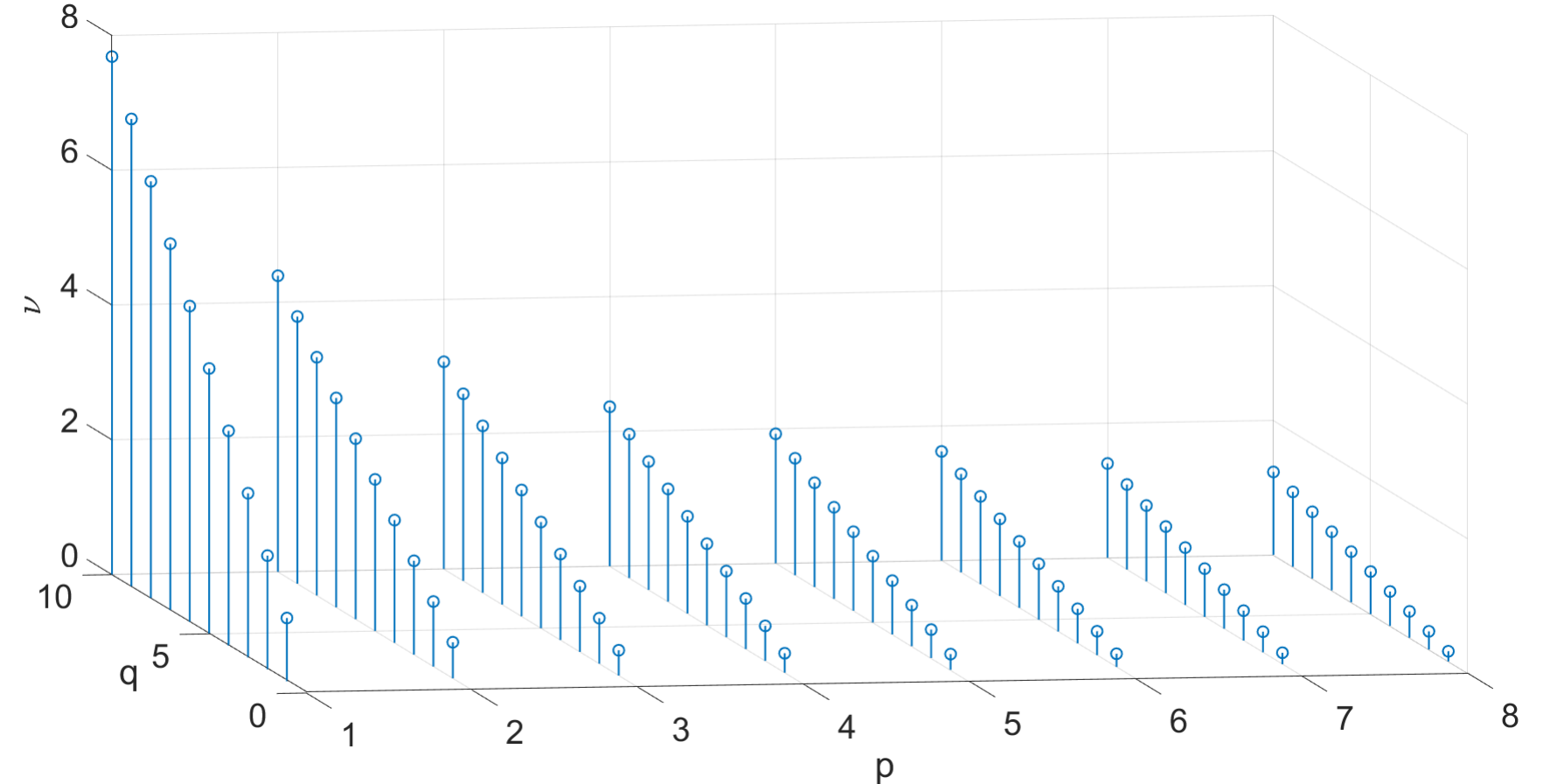}\\
\includegraphics[width=3in]{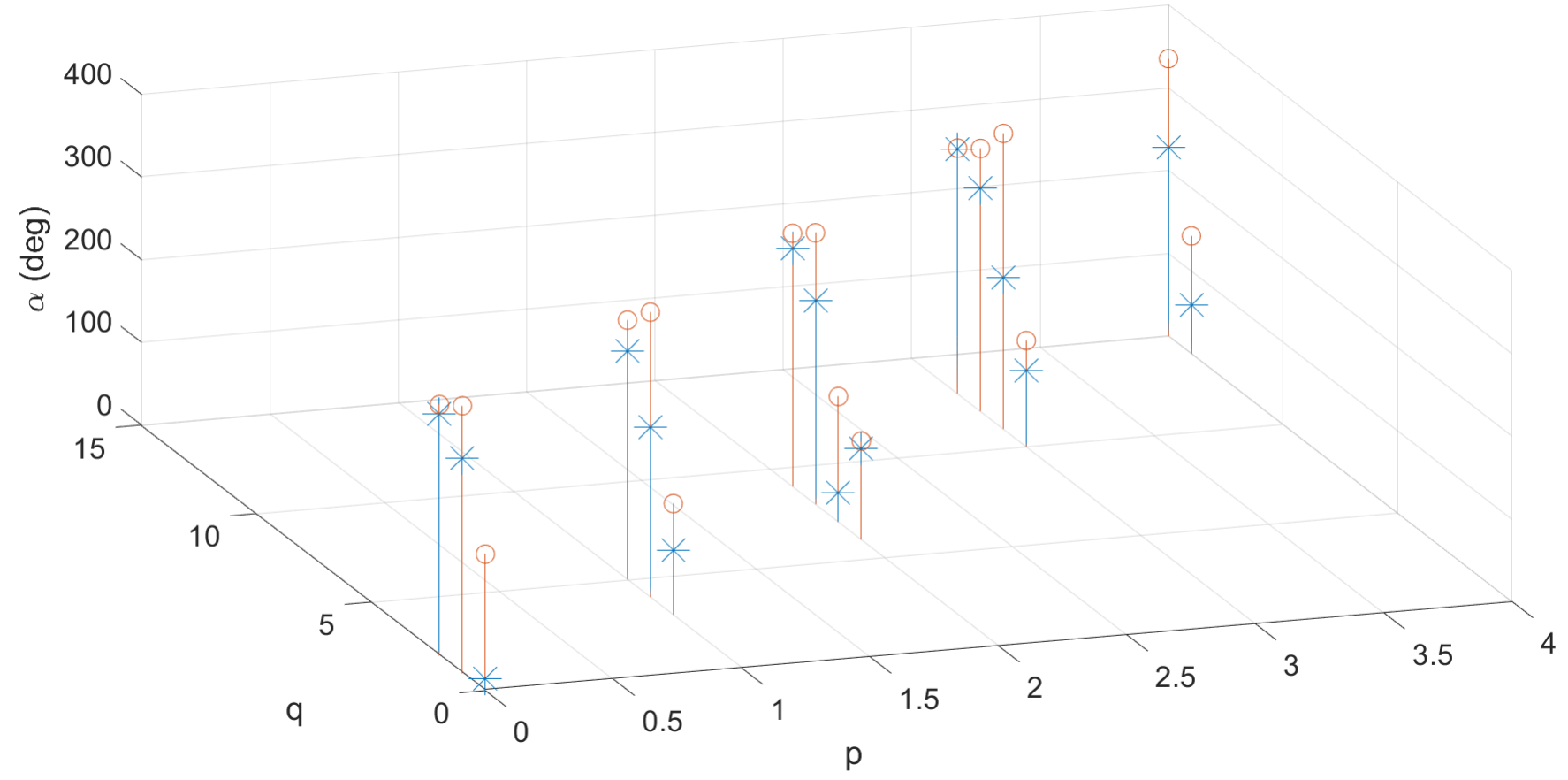}
\includegraphics[width=3in]{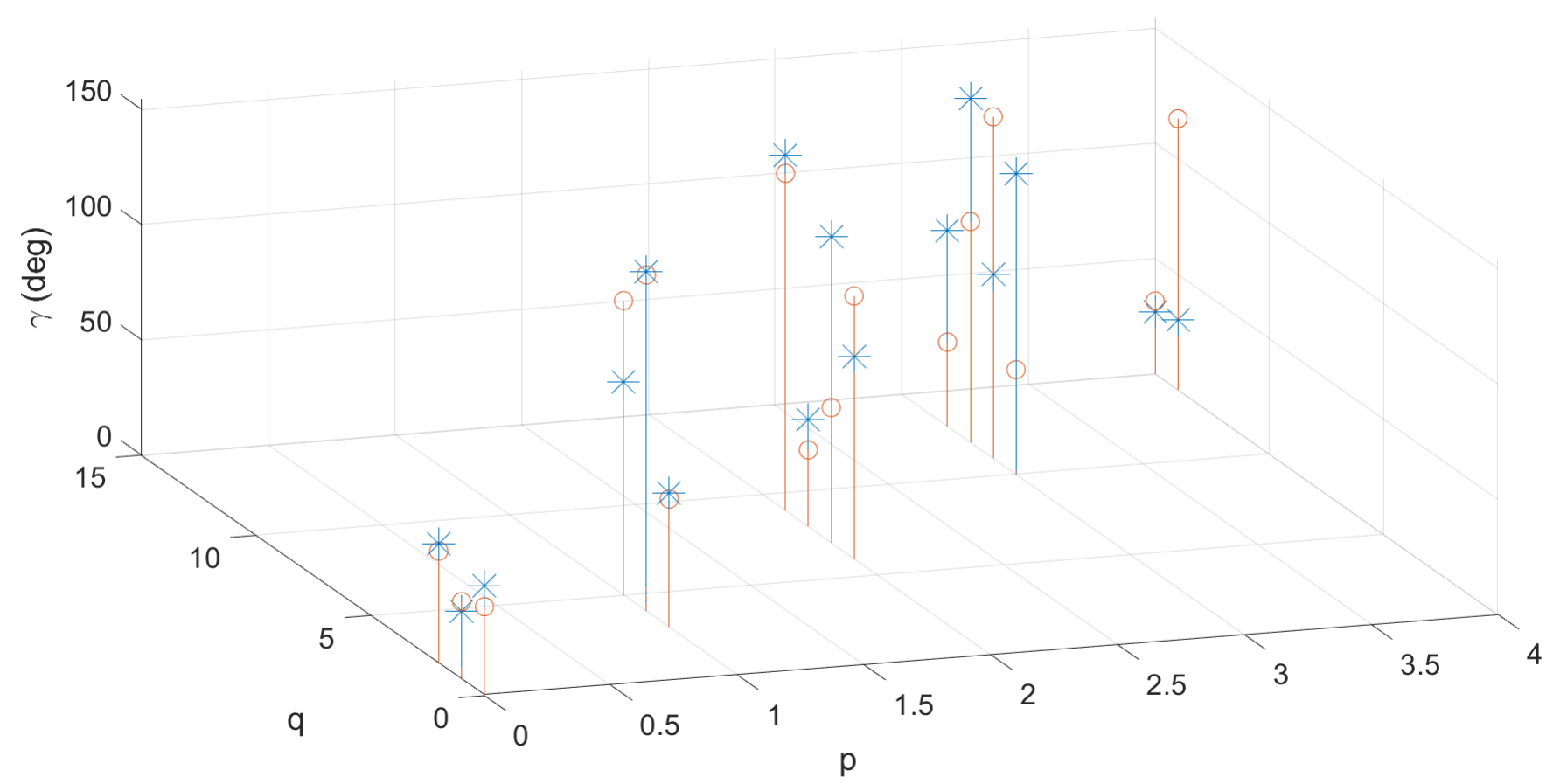}
\caption{(a) $\nu(p,q)$ for the initial condition in Example 1, (b) $\alpha_0(p,q)$ for the initial condition in Example 2 (c) $\gamma(p,q)$ for the initial condition in Example 2}
\label{speed_ratio_fig}
\end{figure}

{\bf Example 1:}
For a given set of initial conditions (in terms of $Z_{\alpha0}$ and $Z_{\beta0}$) and given values of $p$ and $q$, (\ref{speed_ratio2}) provides the speed ratio $\nu$ which will lead to a collision between $A$ and $B$.  Refer Fig \ref{speed_ratio_fig}(a) for an illustration, which is constructed for an example scenario of $Z_\alpha = \frac{\pi}{3}$, $Z_\beta=\frac{\pi}{4}$.  This figure provides a characterization of the speed ratios with which collision will occur for a given $(p,q)$ pair.  

We now look at the use of (\ref{speed_ratio2}) in another context.  For a given speed ratio $\nu$, given initial positions of $A$and $B$ on the surface of the sphere, and a given great circle $C_B$ on which $B$ moves, what is the corresponding great circle $C_A$ of $A$ that will lead to collision at some specified $p,q$?    

{\bf Theorem 2:} Consider two point objects $A$ and $B$, moving with a speed ratio $\nu$.  Assume the great circle on which $B$ moves to be fixed.  Assume the initial positions of $A$ and $B$ on the sphere to be fixed, and such that the initial geodesic distance between $A$ and $B$ is $s_0$, and the angle made by the great circle of $B$ with this initial geodesic is $\beta_0$.   Then, the heading angles of $A$ with which it will be on a collision course with $B$ are given by the values of $\alpha_0$ which are a solution of: 
\begin{eqnarray}
\nu \sin^{-1}\Big[\frac{\sin{\alpha_0} \sin{s_0}}{\sqrt{1 - [\sin \alpha_0 \sin\beta_0 \cos s_0 - \cos \alpha_0 \cos\beta_0]^2}}\Big] 
-\sin^{-1} \Big[\frac{\sin{\beta_0} \sin{s_0}}{\sqrt{1 - [\sin \alpha_0 \sin\beta_0 \cos s_0 - \cos \alpha_0 \cos\beta_0]^2}}\Big] 
- \pi(\nu p-q) = 0   \nonumber \\\label{alpha_eqn}
\end{eqnarray}
{\bf Proof:} To find the heading angles $\alpha_0$ that correspond to a collision, we need to solve (\ref{speed_ratio2}) while treating the quantity $\alpha_0$ as an unknown.    
By performing a re-arrangement of terms in (\ref{speed_ratio2}), we get (\ref{alpha_eqn}).   $\blacksquare$


{\bf Example 2:}
Consider a scenario where the initial conditions are $\beta_0 = \frac{\pi}{3}$, 
$s_0 = \frac{\pi}{6}$ and speed ratio $\nu = 3.6$.  Then, Fig \ref{speed_ratio_fig}(b)  
provdes a characterization of the heading angles $\alpha_0$ of $A$, with which it will intercept $B$, for different $(p,q)$ pairs.  We note that the speed ratio plot of Fig \ref{speed_ratio_fig}(a) depicted a speed ratio that will lead to collision for every $(p,q)$ pair.  However, the plot of Fig \ref{speed_ratio_fig}(b)  
shows that heading angles $\alpha_0$ that will lead to collision occur only for some specific $(p,q)$ pairs, and this is because a solution to (\ref{alpha_eqn}) occurs only for those pairs.  Furthermore, each such $\alpha_0(p,q)$ heading will lead to a different interception angle $\gamma(p,q)$.  
The plot of $\gamma(p,q)$ for this example is given in Fig \ref{speed_ratio_fig}(c).  

%

\section{COLLISION CONDITIONS FOR CIRCULAR PATCHES MOVING ON A SPHERICAL MANIFOLD}

\begin{figure}
\centering
\includegraphics[width=2.5in]{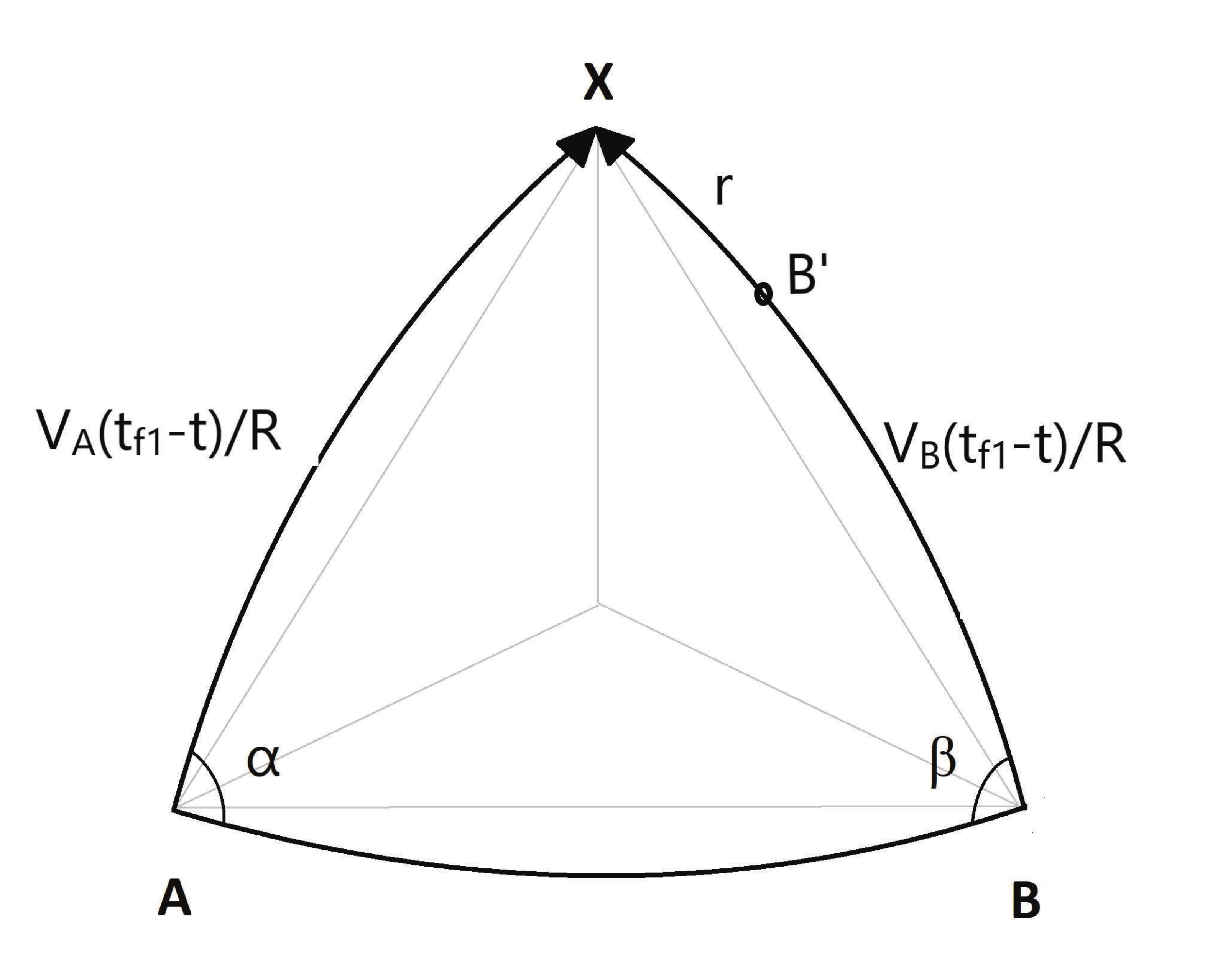}
\includegraphics[width=2.5in]{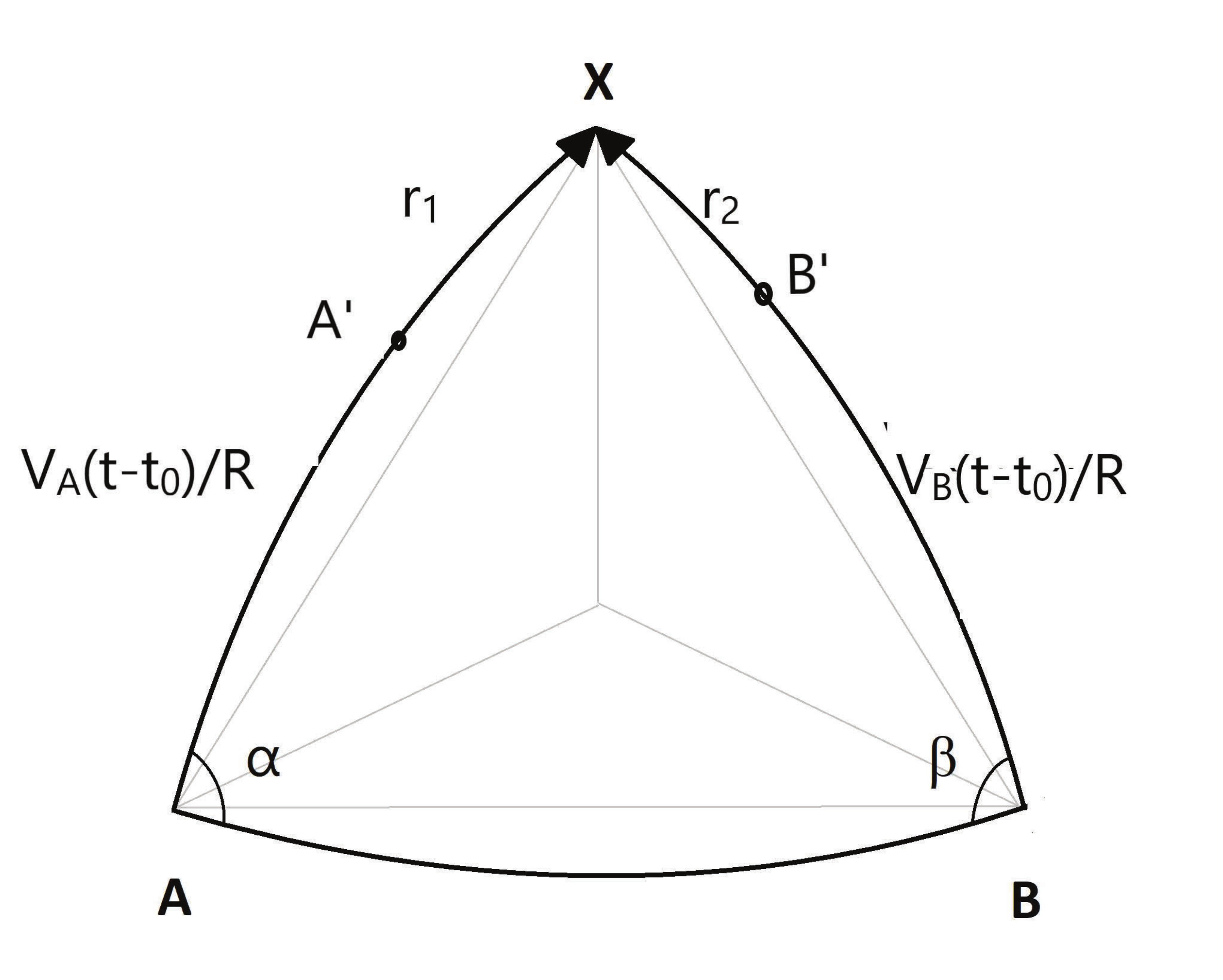}
\caption{(a) Position of $B$ when $A$ has reached a pole, (b) Intermediate positions of $A$ and $B$ before either of them reached a pole}
\label{s_error}
\end{figure}

Eqn (\ref{speed_ratio2}) provides the value of $\nu$ that will cause a collision between $A$ and $B$ at point $X$.  
We now look at the case where such a collision does not occur.  
Refer Fig \ref{s_error}(a), which assumes, without loss of generality, that $A$ reaches $X$ before $B$ does.  Let $t_{f1}$ represent the time at which $A$ reaches $X$, and at that time, let $B$ be located at $B'$, which is a distance $r$ from $X$, as shown.  
Then, applying the sine rule to the triangle, we get the following equation:
\begin{eqnarray}
\frac{\sin (V_A(t_{f1}-t_0)/R)}{\sin \beta} = \frac{\sin ((V_B(t_{f1}-t_0) + r)/R)}{\sin \alpha} 
= \frac{\sin (s_0/R)}{\sin \gamma} \label{sine_rule3}
\end{eqnarray}
Following a set of steps similar to that done earlier, we get the equation:
\begin{equation}
\frac{V_A}{V_B} = \frac{Z_{\beta0}}{Z_{\alpha0} - r/R} \Rightarrow  \frac{r}{R} = \frac{\nu Z_{\alpha0} - Z_{\beta0}}{\nu}  \label{r_error_eqn}
\end{equation} 
Eqn (\ref{r_error_eqn}) thus represents the spherical distance between $A$ and $B$, at the instant when $A$ has reached the pole $X$.  In the case where $B$ has reached the pole $X$ before $A$, we follow a similar series of steps to determine the location of $A$, relative to $X$, as:
\begin{equation}
\frac{r}{R} = \frac{Z_{\beta0} - \nu Z_{\alpha0}}{\nu}  \label{r_error_eqn2}
\end{equation}
Combining (\ref{r_error_eqn}) and (\ref{r_error_eqn}), we define a quantity $r_{mp}$ which we will call as the miss-distance at the pole.  It represents the spherical distance between $A$ and $B$, when either one of them has reached the pole $X$, and is defined as:
\begin{equation}
r_{mp} = \frac{R|Z_{\beta0} - \nu Z_{\alpha0}|}{\nu}  \label{r_mp}
\end{equation}
It is evident from the above that when $\nu$ is equal to the collision speed ratio, that is, $\nu = \frac{Z_{\beta0}}{Z_{\alpha0}}$, then $r_{mp}=0$.   

Refer to Fig \ref{s_error}(b).  Let $A$ and $B$ represent the positions of the two objects at initial time $t=t_0$.  We ask the question: is there an intermediate time $t'$ at which the spherical distance between the two objects is equal to $R_L$, when neither of the objects is necessarily located at a pole.  In the figure, the quantities $r_1(t')$ and $r_2(t')$ represent the distances of the two objects, at time $t'$, from $N$.
Then, applying the spherical sine rule to the triangle $ABN$, we get the following:
\begin{equation}
\frac{\sin (V_A(t'-t_0)/R + r_1(t')/R)}{\sin \beta_0} = \frac{\sin (V_B(t'-t)/R + r_2(t')/R)}{\sin \alpha_0} 
= \frac{\sin (s_0/R)}{\sin \gamma} \label{sine_rule_chk2}
\end{equation}
Considering the first and third terms, followed by the second and third terms, respectively, in the above equation, we get:
\begin{eqnarray}
V_A(t'-t_0)/R &=& \underbrace{\sin^{-1} \Big[\frac{\sin \beta_0 \sin (s_0/R)}{\sin \gamma} \Big]}_{Z_{\beta0}}  - \frac{r_1(t')}{R} \label{VA_eqn_chk3} \\
V_B(t'-t_0)/R &=& \underbrace{\sin^{-1} \Big[\frac{\sin \alpha_0 \sin (s_0/R)}{\sin \gamma} \Big]}_{Z_{\alpha0}} - \frac{r_2(t')}{R} \label{VB_eqn_chk3}
\end{eqnarray}
Dividing (\ref{VA_eqn_chk3}) by (\ref{VB_eqn_chk3}), we get:
\begin{equation}
\nu = \frac{Z_{\beta 0} + q \pi- r_1/R}{Z_{\alpha 0} +p \pi - r_2/R}  \label{r6_eqn1}
\end{equation}
Now, apply the spherical cosine rule to the triangle $A'B'N$, we get:
\begin{equation}
\cos(\frac{R_L}{R}) = \cos(\frac{r_1}{R}) \cos(\frac{r_2}{R}) + \sin(\frac{r_1}{R}) \sin(\frac{r_2}{R}) \cos{\gamma} \label{r6_eqn2}
\end{equation}
In order for the distance between $A$ and $B$ to become less than $R_L$, the above can be written as the following inequality:
\begin{equation}
 \cos(\frac{r_1}{R}) \cos(\frac{r_2}{R}) + \sin(\frac{r_1}{R}) \sin(\frac{r_2}{R}) \cos{\gamma} \geq \cos(\frac{R_L}{R})  \label{r6_eqn3}
\end{equation}

\begin{figure}
\centering
\includegraphics[width=2.5in]{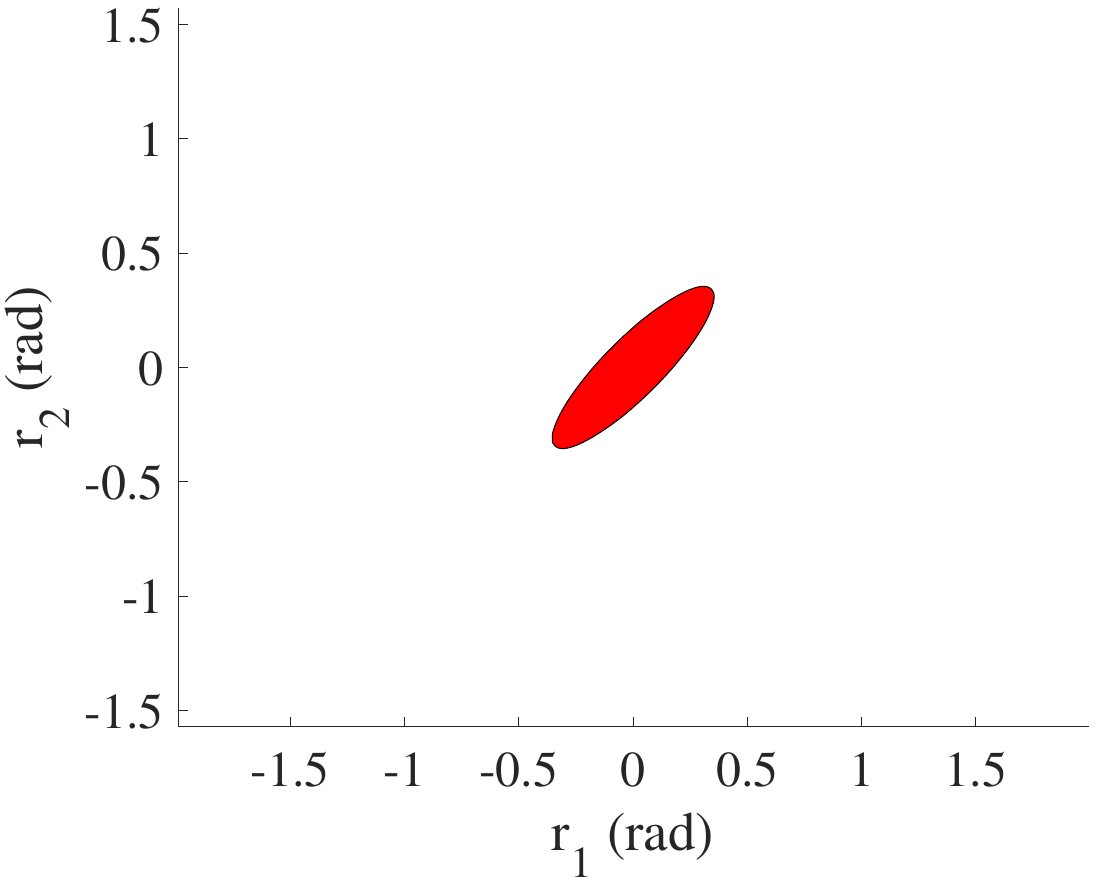}
\includegraphics[width=2.5in]{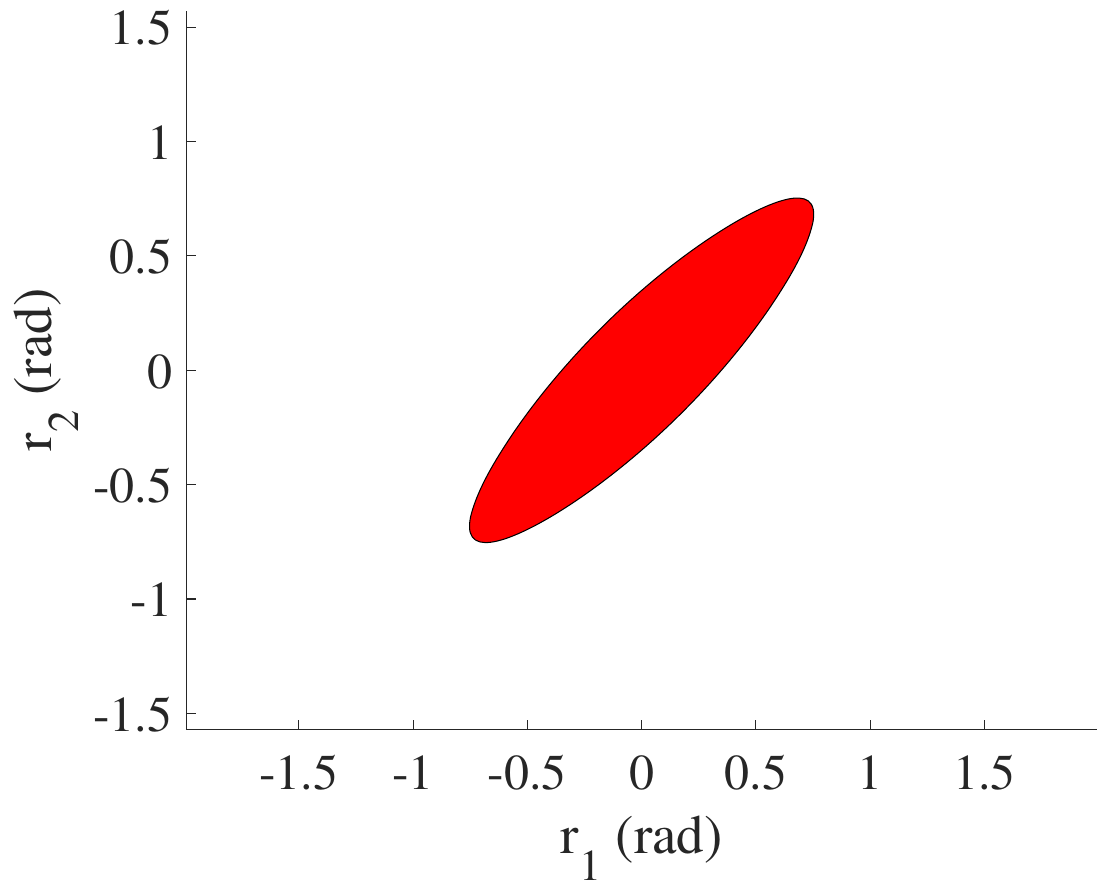}\\
(a) ~~~~~~~~~~~~~~~~~~~~~~~~~~~~~~~~~~~~~~~~~~~~~~~~(b)\\
\includegraphics[width=2.5in]{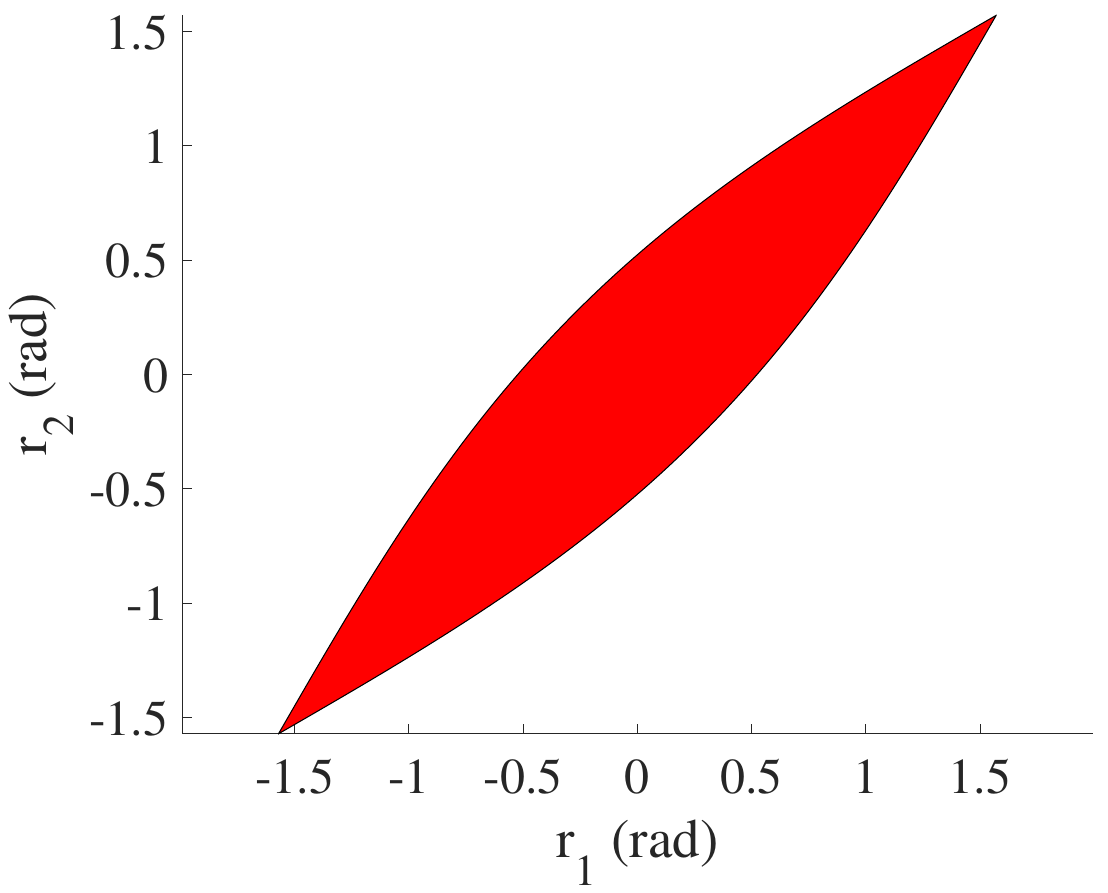}
\includegraphics[width=2.5in]{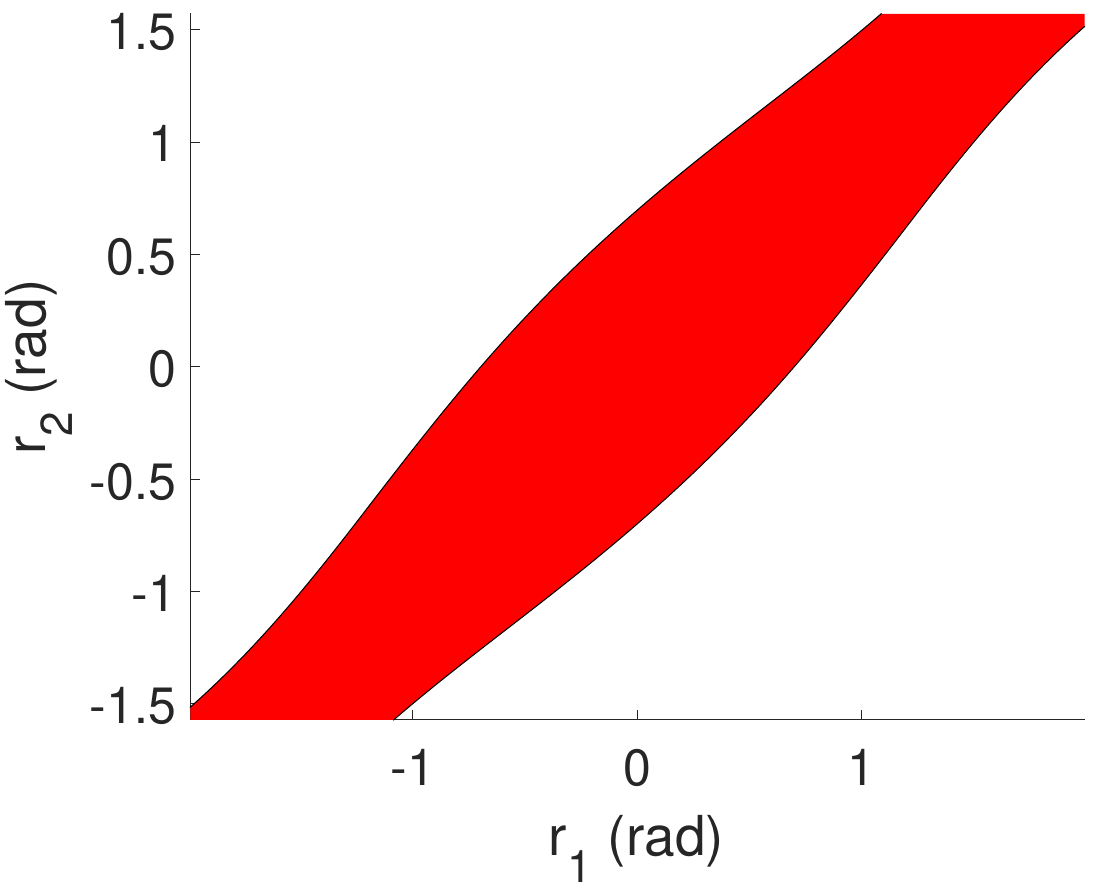}\\
(c) ~~~~~~~~~~~~~~~~~~~~~~~~~~~~~~~~~~~~~~~~~~~~~~~~(d)\\
\caption{Depiction of $\mathcal{C}$ for different values of $R_L$.  (a) $R_L = 10~\rm{deg}$, (b) $R_L = 20~\rm{deg}$, (c) $R_L = 30~\rm{deg}$, (d) $R_L = 40~\rm{deg}$}
\label{Fig_rlethal}
\end{figure}

Let $\mathcal{C}$ represent the region in the $(r_1,r_2)$ space that satisfies the above inequality, for given values of $\gamma$ and $R_L$.  This region $\mathcal{C}$ is shown in Fig \ref{Fig_rlethal}, for different values of $R_L = 10~\rm{deg}, 20~\rm{deg}, 30~\rm{deg}, 40~\rm{deg}$, while keeping $\gamma$ constant at $\gamma = \frac{\pi}{6}$.  As evident from the figure, as $R_L$ increases, the size of $\mathcal{C}$ increases.   
It is possible to compute a closed form solution for the boundary of $\mathcal{C}$, by employing the substitutions: 
\begin{equation}
\cos(\frac{r_2}{R}) = \frac{1-\tan^2 (\frac{r_2}{2R})}{1+\tan^2(\frac{r_2}{2R})}, ~~~ \sin(\frac{r_2}{R}) = \frac{2 \tan (\frac{r_2}{2R})}{1+\tan^2(\frac{r_2}{2R})}
\end{equation}
After making the above substitutions, and for the sake of brevity, employing the notations 
\begin{equation}
c = \tan(\frac{r_2}{2R}), ~~~a = \cos{\gamma}, ~~~b =  \cos(\frac{R_L}{R}) 
\end{equation}
 we can write (\ref{r6_eqn2}) as follows:
\begin{equation}
[b + \cos(\frac{r_1}{R})]c^2 - 2 a \sin(\frac{r_1}{R}) c + [b - a \cos(\frac{r_1}{R})] = 0
\end{equation} 
This can be solved for $c$ and eventually leads to the following solution for $r_2$:
\begin{equation}
\frac{r_2}{R} = 2 \tan^{-1} \Big[\frac{a \sin(\frac{r_1}{R}) \pm \sqrt{a^2 \sin^2(\frac{r_1}{R}) + \cos^2(\frac{r_1}{R}) - b^2}}{b + \cos(\frac{r_1}{R})}     \Big] \label{Cboundary}
\end{equation}
The above equation thus represents a closed-form representation of the boundary of $\mathcal{C}$.  We note from the above that in order for the boundary of $\mathcal{C}$ to be defined, we need the quantity under the square root to be positive.  In other words, the values of $r_1$ that lie on the boundary of $\mathcal{C}$ satisfy:
\begin{equation}
a^2 \sin^2(\frac{r_1}{R}) + \cos^2(\frac{r_1}{R}) \geq b^2
\end{equation}
and the above can be simplified to the following:
\begin{equation}
\sin^2(\frac{r_1}{R}) \leq \frac{\sin^2 (\frac{R_L}{R})}{\sin^2 \gamma} 
\Rightarrow -\sin^{-1}\Big[\frac{\sin(\frac{R_L}{R})}{\sin \gamma}\Big] \leq \frac{r_1}{R} \leq \sin^{-1}\Big[\frac{\sin(\frac{R_L}{R})}{\sin \gamma}   \Big] \label{r1_boundary}
\end{equation}
From the above, it is evident that when $\frac{R_L}{R} \geq \gamma$, then the values of $r_1$ that belong to the boundary of $\mathcal{C}$ will span the full range of $[-\pi/2,\pi/2]$ as shown in Fig \ref{Fig_rlethal}(c), (d).  For the more phsically meaningful scenarios of $\frac{R_L}{R} < \gamma$, this is obviously not the case, as is depicted in Fig \ref{Fig_rlethal}(a),(b).

\begin{figure}
\centering
\includegraphics[width=3.2in]{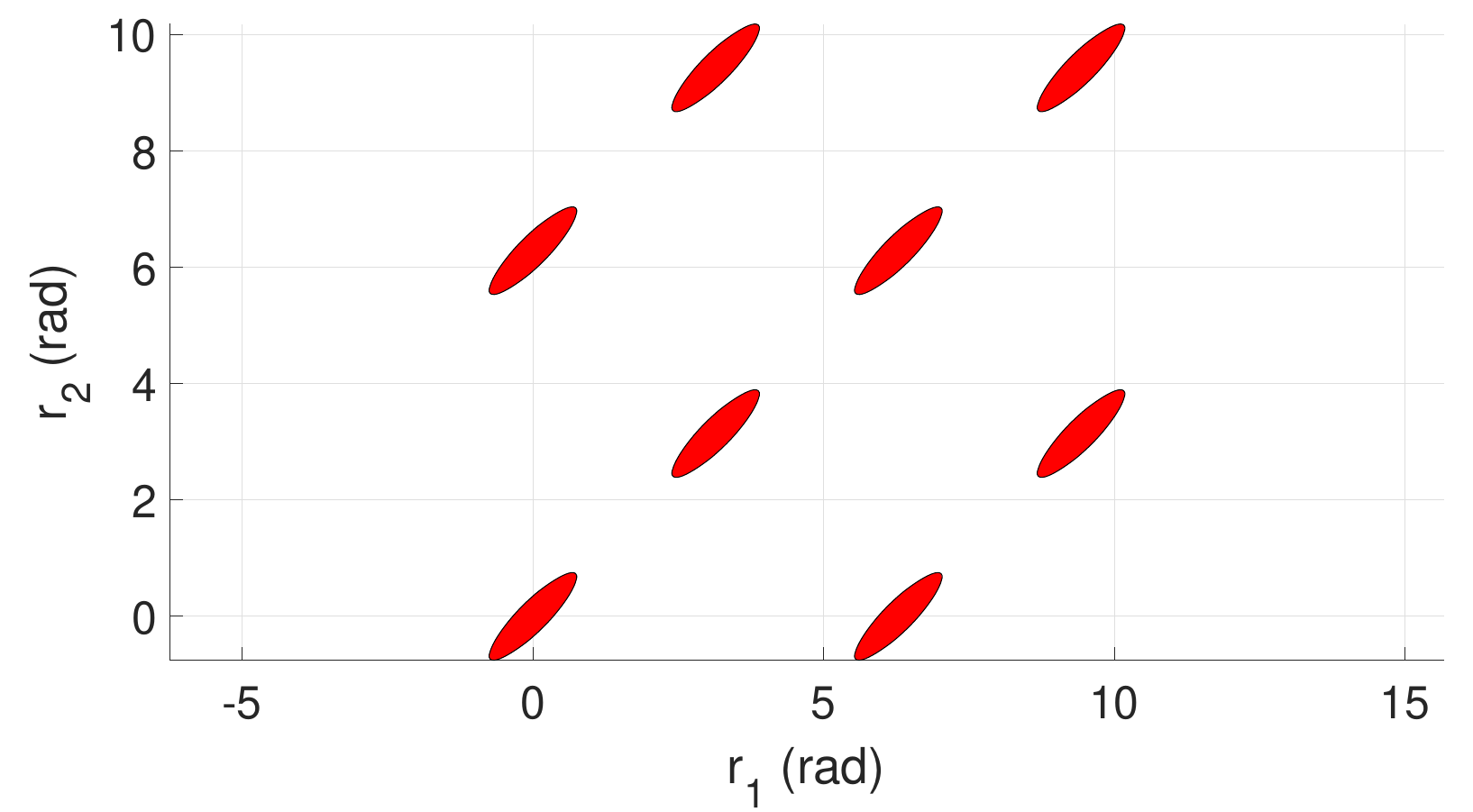}
\includegraphics[width=2.2in]{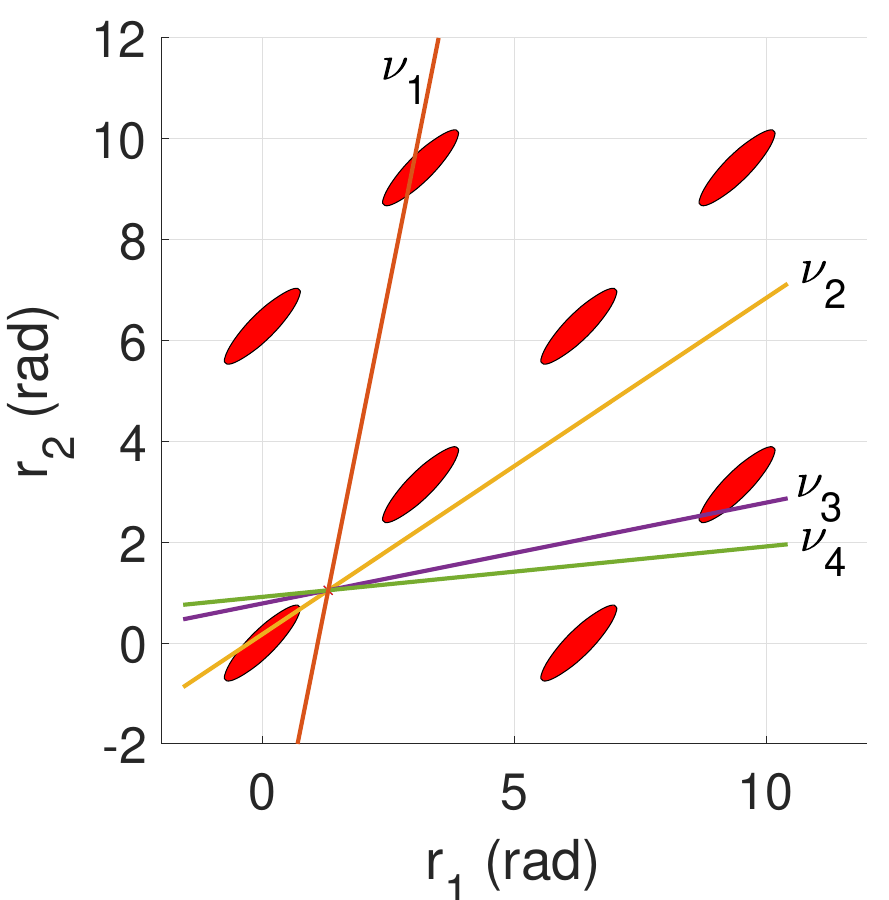}\\
(a) ~~~~~~~~~~~~~~~~~~~~~~~~~~~~~(b)\\
\caption{(a) Depiction of ${\mathcal{C}}_{0.0}$, ${\mathcal{C}}_{2\pi,0}$, ${\mathcal{C}}_{\pi,\pi}$, ${\mathcal{C}}_{3\pi,\pi}$, ${\mathcal{C}}_{0,2\pi}$, ${\mathcal{C}}_{2\pi,2\pi}$, ${\mathcal{C}}_{\pi,3\pi}$, ${\mathcal{C}}_{3\pi,3\pi}$, (b) Influence of different speed ratios: $\nu_1=0.2, \nu_2=1.5,\nu_3=5,\nu_4=10$}
\label{Fig_rlethal_grid}
\end{figure}

Fig \ref{Fig_rlethal} shows $\mathcal{C}$ around one of the poles (say the North Pole $N$).  It represents the set of spherical distance pairs $(r_1,r_2)$ of $A$ and $B$, respectively, from $N$, for which the spherical distance between $A$ and $B$ is less than or equal to $R_L$.  Let us refer to this set as ${\mathcal{C}_{0,0}}$.  We can determine the corresponding contour around the South Pole $S$, by simply adding $\pi$ radians to the values of $r_1$ that lie on the boundary of $\mathcal{C}_{0,0}$ (as determined from (\ref{r1_boundary})) and finding the corresponding values of $r_2$, using (\ref{Cboundary}).  This leads to a region, which we will refer as ${\mathcal{C}}_{\pi,\pi}$.  In a similar fashion, we can find regions ${\mathcal{C}}_{2\pi,0}$, ${\mathcal{C}}_{2\pi,2\pi}$, and so on.  In general, we can compute the regions ${\mathcal{C}}_{i,j}$, as shown in Fig \ref{Fig_rlethal_grid}(a).  This plot is shown for the values $R_L=20~\rm{deg}$, $\gamma = 30~\rm{deg}$.

Note that (\ref{r6_eqn1}) can be written explicitly in terms of $r_2$ as follows:
\begin{equation}
\frac{r_2}{R} = (\frac{1}{\nu}) (\frac{r_1}{R}) - \frac{Z_{\beta 0}}{\nu} + Z_{\alpha0} + \pi (p-\frac{q}{\nu}) \label{r6_eqn4} 
\end{equation}
and this represents a line in $(r_1,r_2)$ space.  The slope of this line varies with the speed ratio $\nu$ and is shown in Fig \ref{Fig_rlethal_grid}(b) for different values of $\nu$, and for a single initial condition $(Z_{\beta 0},Z_{\alpha 0})$.  If, for a given value of $\nu$, this line intersects one of the ${\mathcal{C}}_{i,j}$ contours then this indicates a collision between the two objects for that $(i,j)$ pair.  For example, Fig \ref{Fig_rlethal_grid}(b)   
indicates that with the speed ratio $\nu_1$, the $(r_1,r_2)$ line intersects the ${\mathcal{C}}_{\pi,3\pi}$ which means that a collision will occur when the objects are near the South Pole $S$, and this collision will occur during $A's$ current revolution and after $B$ has completed one revolution.  For the speed ratio $\nu_4$, the $(r_1,r_2)$ line does not intersect any of the ${\mathcal{C}}_{i,j}$ contours shown in the figure, and this means that there will be no collisions occuring for the $(i,j)$ range shown in this figure.
    

The prediction of such a collision can be mathematically performed as follows.  Substitute (\ref{r6_eqn4}) in (\ref{r6_eqn2}) to get the following equation which is implicit in a single unknown $r_1$.
\begin{equation}
\cos(\frac{r_1}{R}) \cos(\Big[(\frac{1}{\nu}) (\frac{r_1}{R}) - \frac{Z_{\beta 0}}{\nu} + Z_{\alpha0}\Big]) 
+ \sin(\frac{r_1}{R}) \sin(\Big[(\frac{1}{\nu}) (\frac{r_1}{R}) - \frac{Z_{\beta 0}}{\nu} + Z_{\alpha0}\Big]) \cos{\gamma} 
-   \cos(\frac{R_L}{R})  = 0 \label{r6_eqn3}
\end{equation}
For a given initial condition, speed ratio and lethal radius, if the above equation admits no solution in $r_1$, then collision will never occur.  If it does admit a solution, then collision will occur.

\begin{figure}
\centering
\includegraphics[width=3in]{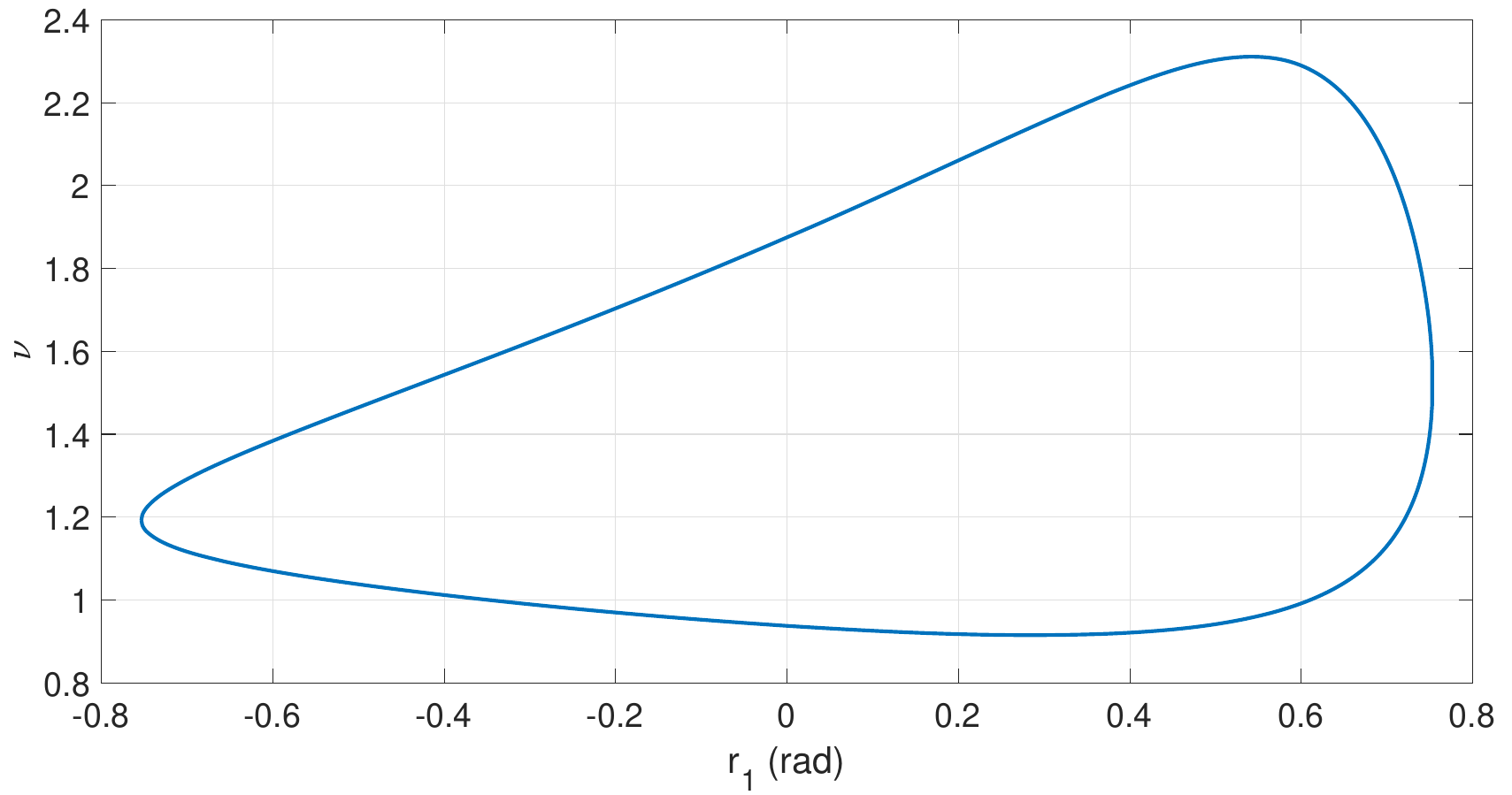}
\includegraphics[width=3in]{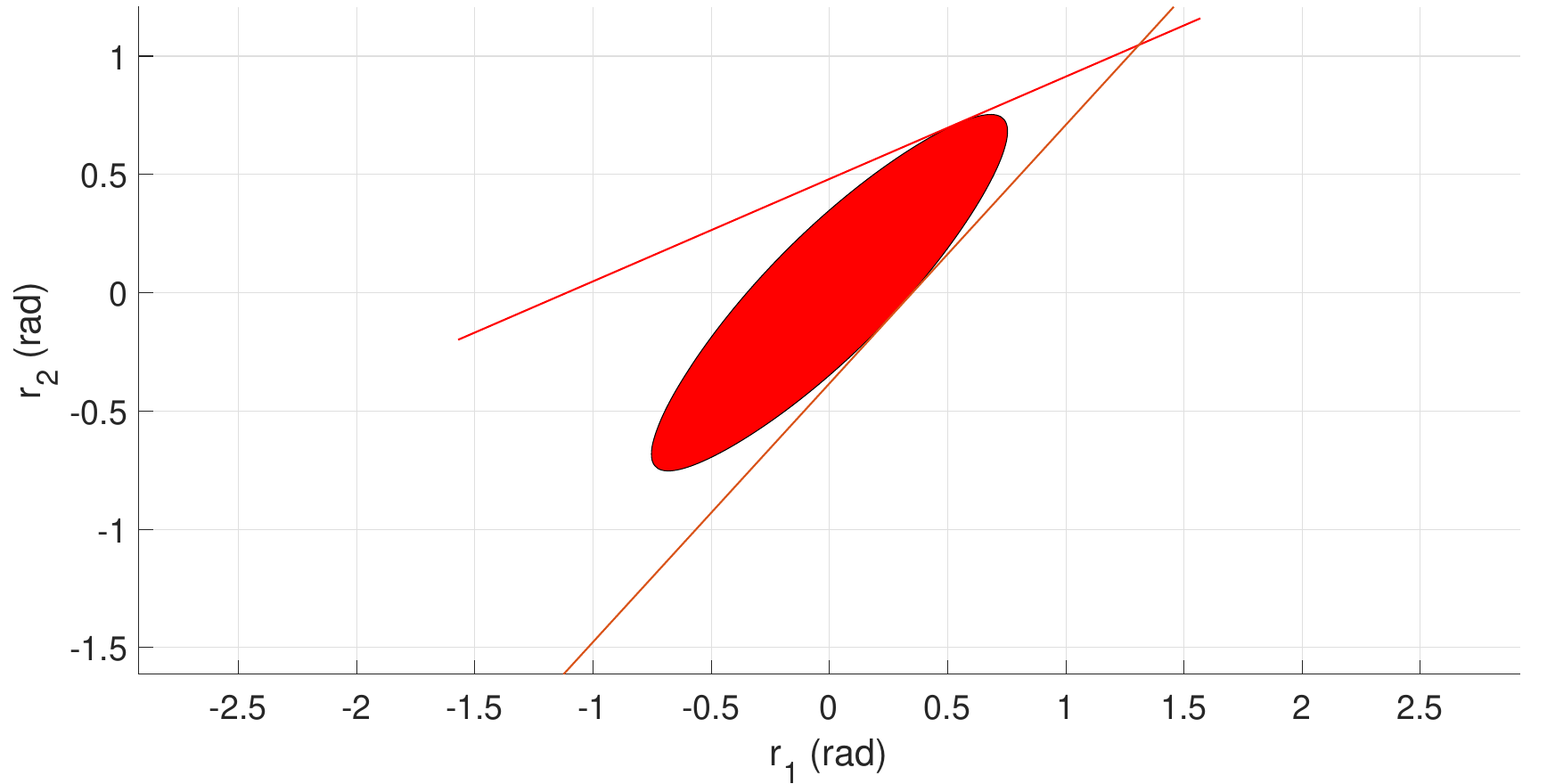}\\
(a)~~~~~~~~~~~~~~~~~~~~~~~~~~~~~~~~~~~~~~~~~(b)\\
\caption{Illustration of (a) Eqn (\ref{nu_ratio}) for different values of $\nu$, (b) speed ratio limits as tangents to a $\mathcal{C}$ contour}
\label{Fig_nu_ratio1}
\end{figure}

From (\ref{r6_eqn3}), we can determine the values of $\nu$ for which a double root (that is, repeated root) in $r_1$ exists.  With reference to Fig \ref{Fig_rlethal_grid}, this corresponds to the phenomenon of the $(r_1,r_2)$ line being tangent to one of the ${\mathcal{C}}_{i,j}$ contours.  By identifying these individual values of $\nu$, we can subsequently determine the range of $\nu$ with which collision will occur near a specified pole, and after a pre-specified number of revolutions of $A$ and $B$.   
Substituting $r_2$ from (\ref{r6_eqn4}) in (\ref{Cboundary}), we get the following: 
\begin{equation}
(\frac{1}{\nu}) (\frac{r_1}{R}) - \frac{Z_{\beta 0}}{\nu} + Z_{\alpha0} + \pi(p-\frac{q}{\nu}) = 2 \tan^{-1} \Big[\frac{a \sin(\frac{r_1}{R}) \pm \sqrt{a^2 \sin^2(\frac{r_1}{R}) + \cos^2(\frac{r_1}{R}) - b^2}}{b + \cos(\frac{r_1}{R})}     \Big]
\end{equation}
The above equation can be explicitly solved for $\nu$ to obtain the following:
\begin{equation}
\nu = \frac{(\frac{r_1}{R}-Z_{\beta0} - q\pi)}{2 \tan^{-1} \Big[\frac{a \sin(\frac{r_1}{R}) \pm \sqrt{a^2 \sin^2(\frac{r_1}{R}) + \cos^2(\frac{r_1}{R}) - b^2}}{b + \cos(\frac{r_1}{R})}     \Big] - Z_{\alpha0} - p\pi} \label{nu_ratio}
\end{equation}
This equation is depicted in Fig \ref{Fig_nu_ratio1}(a), and shows, for each value of $r_1$, the range of speed ratios with which $A$ will intercept $B$.  The maximum and minimum values of $\nu$ in this figure represent the upper and lower limits of the speed ratio with which $A$ will intercept $B$.  For the example shown in this figure, $A$ will intercept $B$ for any speed ratio which lies in the range $[0.915, 2.311]$, and the upper and lower limits correspond to the slopes of the lines drawn from the initial condition on the $(r_1,r_2)$ plane that are tangent to the $\mathcal{C}$ contour as shown in Fig \ref{Fig_nu_ratio1}(b).   
Note that this interception will occur at $N$ and in the course of the current revolution of $A$ and $B$.

\begin{figure}
\centering
\includegraphics[width=3in]{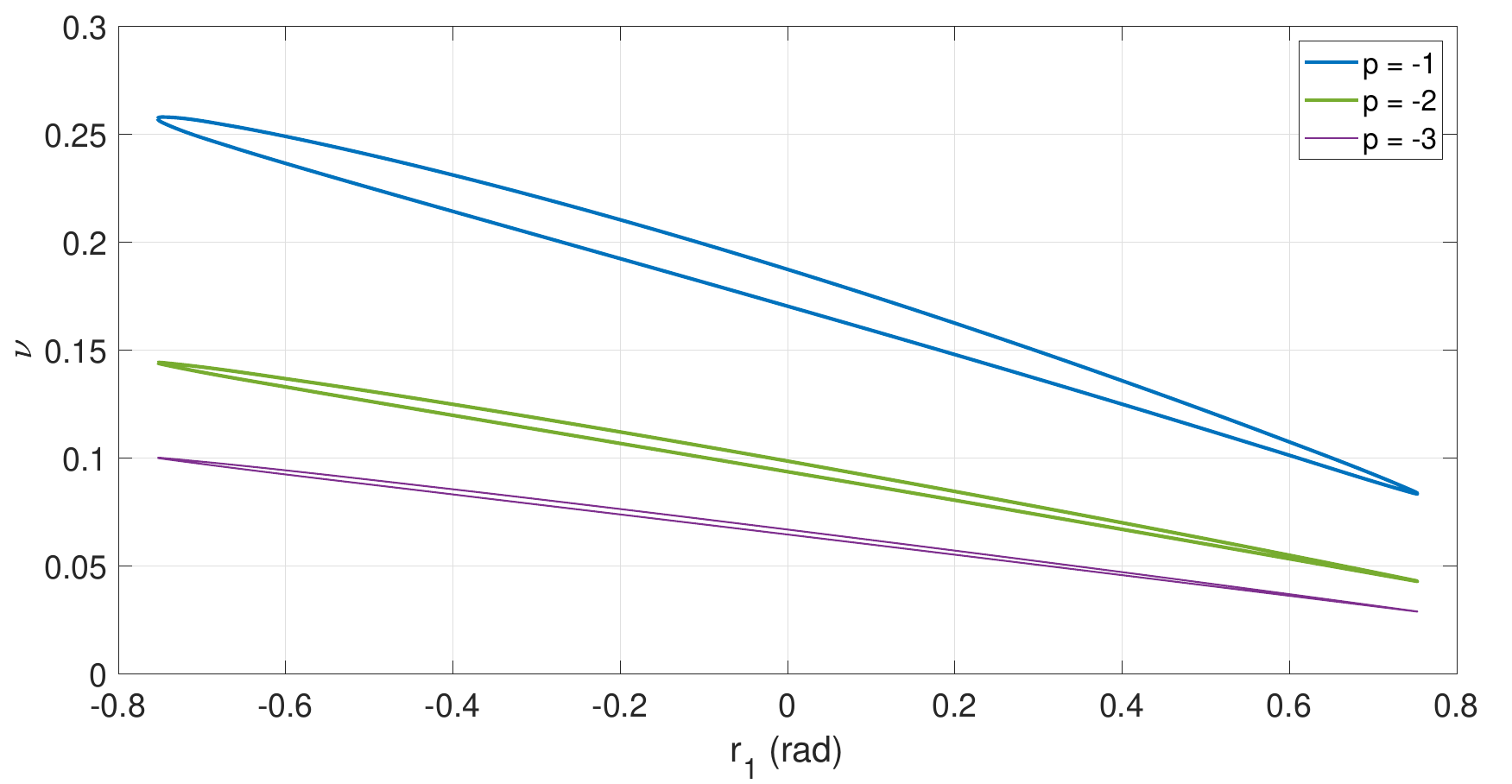}
\includegraphics[width=3in]{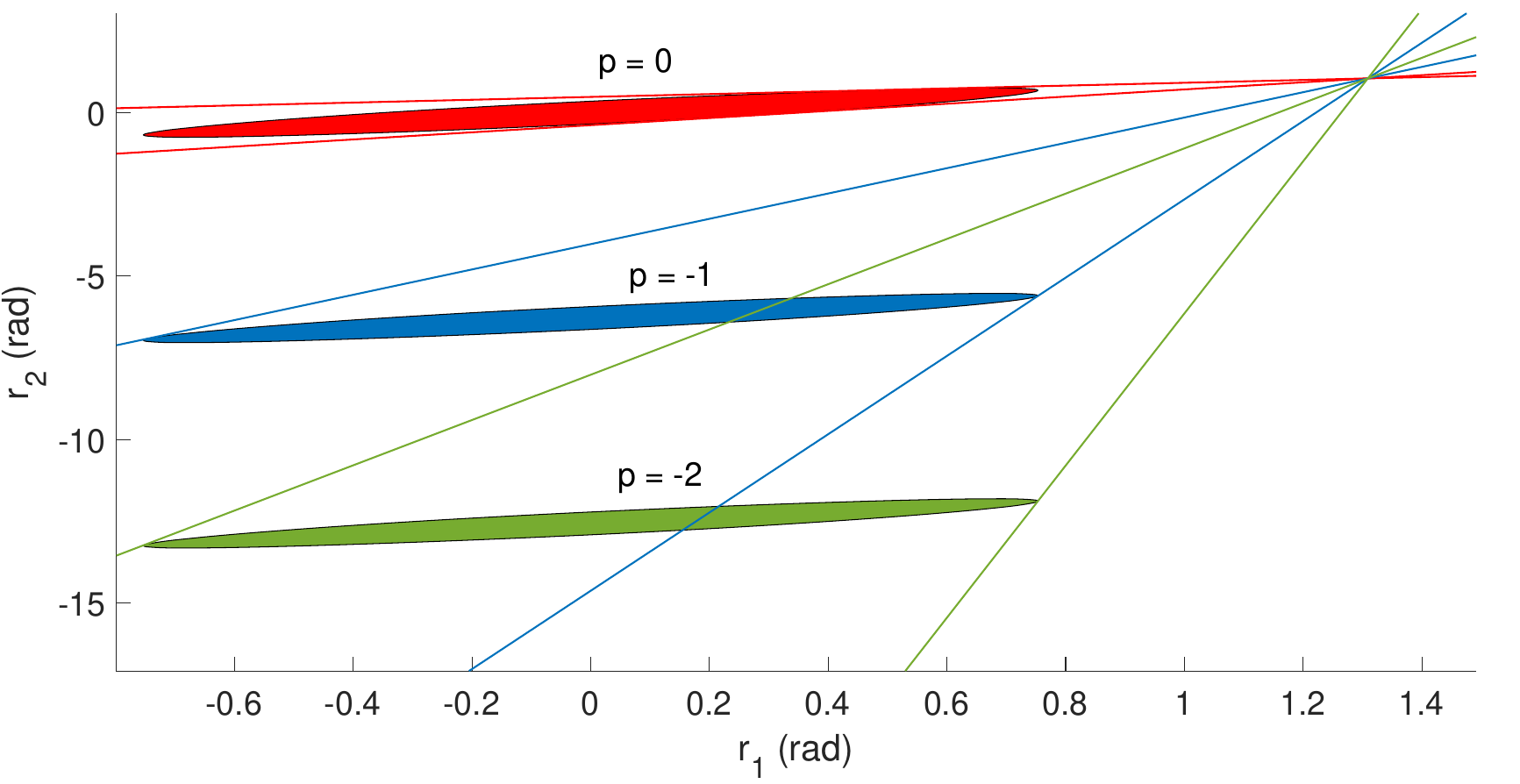}\\
(a)~~~~~~~~~~~~~~~~~~~~~~~~~~~~~~~~~~~~~~~~~(b)\\
\caption{Illustrations of (a) Eqn (\ref{nu_ratio}) for $q=0$ and different values of $p$, (b) speed ratio limits as tangents to several $\mathcal{C}$ contours}
\label{Fig_nu_ratio3}
\end{figure}

Eqn (\ref{nu_ratio}) is depicted for $q=0$ and different values of  $p=-1,-2,-3$ in Fig \ref{Fig_nu_ratio3}(a).    
From this figure, we see that for $p=-1$, interception will occur if the speed ratio lies in the range $[0.0835, 0.2582]$.  Similarly, for $p=-2$ and $p=-3$, interception will occur if the speed ratio lies in the ranges $[0.0430,0.1444]$ and $[0.0289,0.1003]$, respectively.  The upper and lower limits in each of these ranges correspond to the slopes of the lines that are tangent to the corresponding $\mathcal{C}$ contour in the $(r_1,r_2)$ plane, as is depicted in Fig \ref{Fig_nu_ratio3}(b).   
We now state the following theorem:

{\bf Theorem 3:} Consider a point object $A$ and a circular object $B$ moving on the surface of a sphere along great circles that are on planes with a relative angle $\gamma$.  Let the radius of $B$ be $R_L$.  Then, $A$ and $B$ are on a collision course if their speed ratio $\nu$ is such that $\nu \in [\nu_{min},\nu_{max}]$, where for each $(p,q)$ pair, $\nu_{min}$ and $\nu_{max}$ are defined as follows:
\begin{eqnarray}
\centering
\nu_{min} &=& \min_{r_1 \in r_{1C}} \displaystyle\frac{(\frac{r_1}{R}-Z_{\beta0} - q\pi)}{2 \tan^{-1} \Big[\frac{a \sin(\frac{r_1}{R}) \pm \sqrt{a^2 \sin^2(\frac{r_1}{R}) + \cos^2(\frac{r_1}{R}) - b^2}}{b + \cos(\frac{r_1}{R})}     \Big] - Z_{\alpha0} - p\pi} \nonumber \\
\nu_{max} &=& \max_{r_1 \in r_{1C}} \displaystyle\frac{(\frac{r_1}{R}-Z_{\beta0} - q\pi)}{2 \tan^{-1} \Big[\frac{a \sin(\frac{r_1}{R}) \pm \sqrt{a^2 \sin^2(\frac{r_1}{R}) + \cos^2(\frac{r_1}{R}) - b^2}}{b + \cos(\frac{r_1}{R})}     \Big] - Z_{\alpha0} - p\pi} , ~~p,q = 0,1,\cdots 
\end{eqnarray}
{\bf Proof:} Follows from the above analysis.   $\blacksquare$

For a given speed ratio, the cone of directions that will lead to collision is computed as follows.
The equation of the tangent to the $\mathcal{C}$ contour is found by differentiation to be:
\begin{equation}
\frac{dr_2}{dr_1} = \frac{\sin(r_1/R) \cos(r_2/R) - \cos(r_1/R) \sin(r_2/R) \cos{\gamma}}{\sin(r_1/R) \cos(r_2/R) \cos{\gamma} - \cos(r_1/R) \sin(r_2/R)}
\end{equation}
Note that the above equation is a function of $\gamma$.  We can then find the specific $(r_1,r_2)$ pairs for which the above slope is equal to the speed ratio $\nu$.  Equating the right hand side of the above equation to $\frac{1}{\nu}$, we get the following:
\begin{equation}
\frac{\tan(r_2/R)}{\tan(r_1/R)} = \underbrace{\frac{\nu-\cos{\gamma}}{\nu-1}}_{M} 
\Rightarrow \tan(r_2/R) = M \tan(r_1/R)
\end{equation}
where we note that the quantity $M$ always satisfies $M \geq 1$, and $M=1$ only occurs for the special case when $\gamma = \frac{\pi}{2}$.   From the above, we get 
\begin{equation}
\cos(r_2/R) = \frac{\cos(r_1/R)}{\sqrt{M^2 \sin^2(r_1/R) + \cos^2(r_1/R)}}, ~~~~~~~~
\sin(r_2/R) = \frac{M\sin(r_1/R)}{\sqrt{M^2 \sin^2(r_1/R) + \cos^2(r_1/R)}}
\end{equation}
These can be substituted back into \ref{r6_eqn3} (now used with the equality sign) to arrive an equation in a single unknown $(r_1/R)$ as follows:
\begin{equation}
\cos^2(r_1/R) + M \sin^2(r_1/R) 
= \cos(R_L/R) \sqrt{M^2 \sin^2(r_1/R) + \cos^2(r_1/R)}
\end{equation}
In the above equation, make the substitution $Z = \cos^2(r_1/R)$ and then after some algebraic manipulations, we get the following quartic equation in $Z$:
\begin{equation}
(M-1)^2 Z^4 - (M-1) [2M - (1+M) \cos^2(R_L/R)] Z^2 
+ M^2 \sin^2(R_L/R) = 0 
\end{equation}  
It can be shown that two values of $Z$ are always complex, and the two real values are given by:
\begin{equation}
Z = \Bigg(\frac{[2M - (1+M) \cos^2(\frac{R_L}{R})] + \cos(\frac{R_L}{R}) \sqrt{(1+M)^2 \cos^2(\frac{R_L}{R}) - 4M}}{2(M-1)}\Bigg)^{1/2} \nonumber
\end{equation}
From the above equation, we can find the two values of $r_1$ at which the line of slope $\nu$ is tangential to $\mathcal{C}$ and these can be then used to find the corresponding values of $r_2$ as well.  Denote these values as $(\bar{r}_1,\bar{r}_2)$.  
We can now write the equation of the line having slope $\nu$ and that passes through $(\bar{r}_1,\bar{r}_2)$.  We note that there will be two such lines, since there are two pairs of values of $(\bar{r}_1,\bar{r}_2)$.  This equation will have the form:
\begin{equation}
\frac{r_2}{R} = \frac{1}{\nu} \frac{r_1}{R} - \frac{\bar{r}_1}{\nu} + \bar{r}_2  \label{eqn_tangent}
\end{equation}
Any initial conditions that lie on this line will lead to a collision.  Note that in the above equation, the quantities $\bar{r}_1,\bar{r}_2$ are functions of $\gamma$.  We note that $\frac{r_1}{R}$ represents the distance of $A$ from the collision pole, and this is the same as $Z_{\beta 0} + q \pi$, and $\frac{r_2}{R}$ represents the distance of $B$ from the collision pole, which is the same as $Z_{\alpha0}+ p \pi$.  In other words,
\begin{eqnarray}
\frac{r_1}{R} &=& \sin^{-1} \Big[\frac{\sin{\beta_0} \sin(s_0/R)}{\sin \gamma} \Big] + q\pi \nonumber \\
\frac{r_2}{R} &=& \sin^{-1} \Big[\frac{\sin{\alpha_0} \sin(s_0/R)}{\sin \gamma} \Big] + p\pi
\end{eqnarray}
Substituting the above into (\ref{eqn_tangent}), we get:
\begin{equation}
\sin^{-1} \Big[\frac{\sin{\alpha_0} \sin(s_0/R)}{\sin \gamma} \Big] + p\pi     
= \frac{1}{\nu} \sin^{-1} \Big[\frac{\sin{\beta_0} \sin(s_0/R)}{\sin \gamma} \Big] + \frac{q}{\nu}\pi- \frac{\bar{r}_1}{\nu} + \bar{r}_2 
\end{equation}
This leads us to the following theorem.

{\bf Theorem 4:} Consider a point object $A$ and a circular object $B$ of radius $R_L$, moving with a speed ratio $\nu$.  Assume the great circle on which $B$ moves to be fixed.  Assume the initial positions of $A$ and $B$ on the sphere to be fixed, and such that the initial geodesic distance between $A$ and $B$ is $s_0$, and the angle made by the great circle of $B$ with this initial geodesic is $\beta_0$.   Then, $A$ will be on a collision course with $B$ if its heading angle $\alpha_0$ lies in the collision cone, where the boundaries of the collision cone are determined as solutions $\alpha_0$ of the following equation:
\begin{equation}
\sin^{-1} \Big[\frac{\sin{\alpha_0} \sin(s_0/R)}{\sin \gamma} \Big] + p \pi   
= \frac{1}{\nu} \sin^{-1} \Big[\frac{\sin{\beta_0} \sin(s_0/R)}{\sin \gamma} \Big] + q\pi - \frac{\bar{r}_1}{\nu} + \bar{r}_2, ~~p,q = 0,1,\cdots 
\end{equation}
{\bf Proof:} Follows from the above analysis.  $\blacksquare$

\begin{figure}
\centering
\includegraphics[width=2.8in]{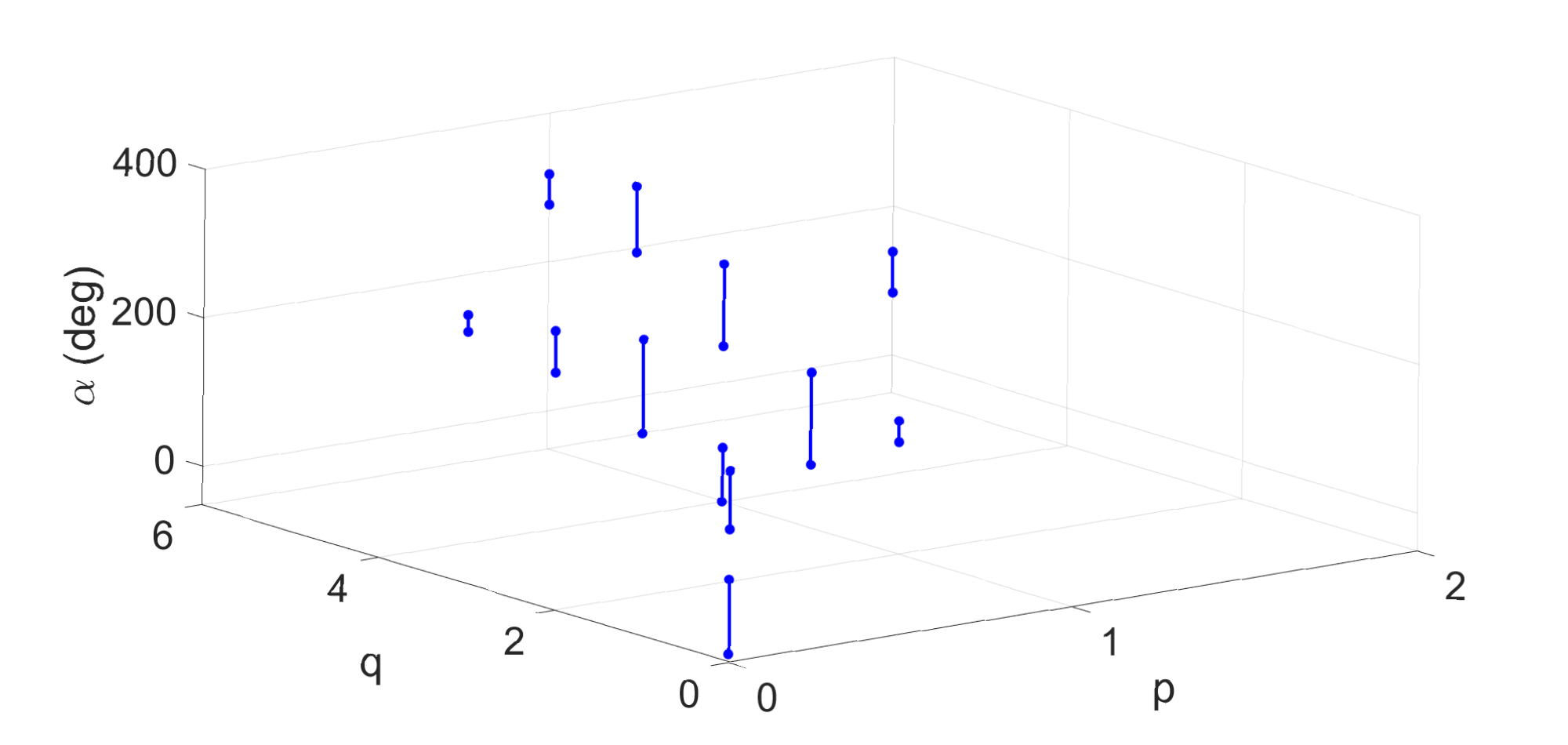}
\includegraphics[width=2.8in]{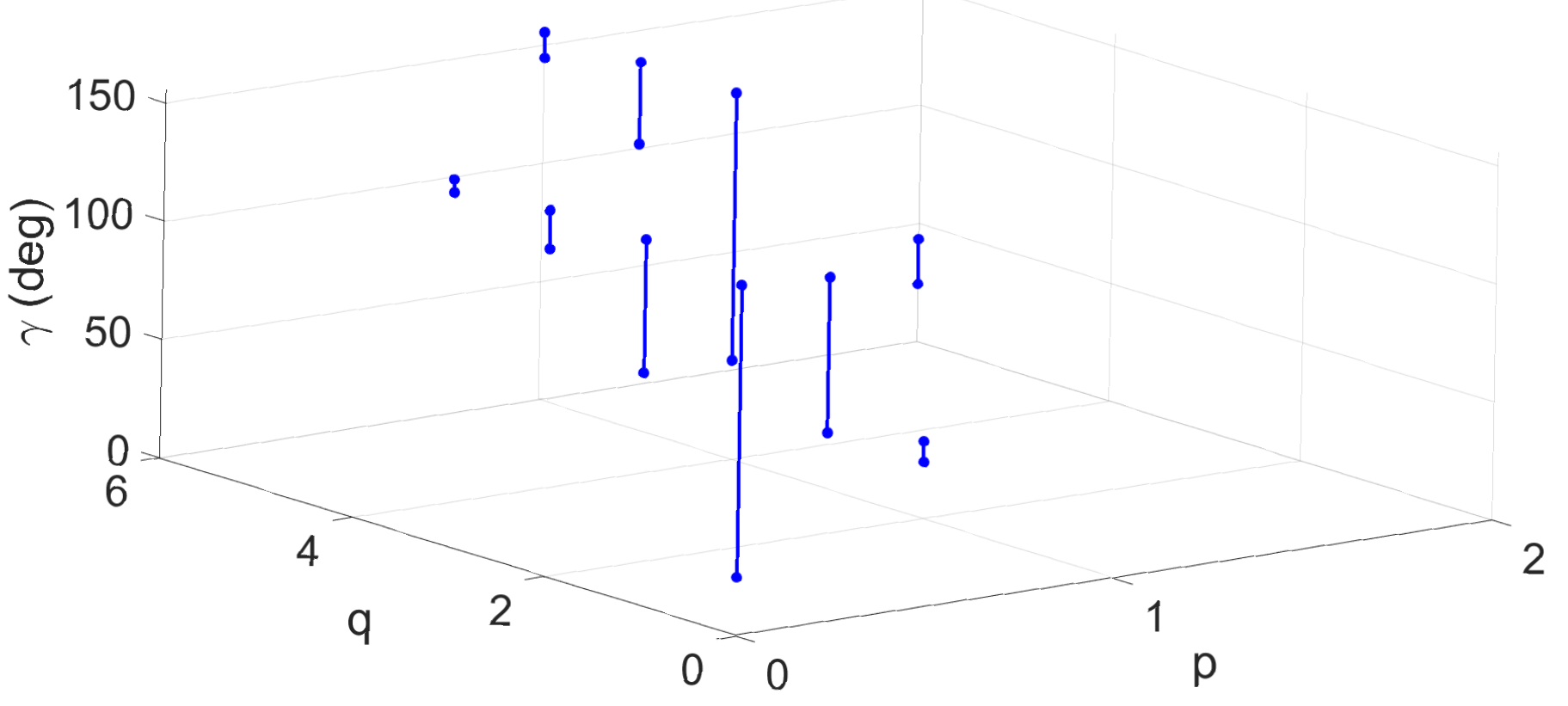}
\caption{Depiction of (a) $\alpha$ limits and (b) $\gamma$  for different $(p,q)$ combinations for the initial conditions in Example 3}
\label{Fig_alpha_cone}
\end{figure}

{\bf Example 3:} Consider a scenario with the speed ratio $\nu=3.6$, and initial conditions $s_0=\frac{\pi}{6}$, $\beta_0 = \frac{\pi}{3}$, $R=\frac{\pi}{9}$.  Fig \ref{Fig_alpha_cone}(a) illustrates the collision cone as computed for different $(p,q)$ combinations, and Fig \ref{Fig_alpha_cone}(b)   
provides the range of $\gamma$ which is achievable for each of these $(p,q)$ values.  
%

\section{CONCLUSIONS}
In this paper, we address the problem of determining analytical conditions to predict the occurrence of collision between objects moving on a spherical manifold.  We first develop these conditions for the scenario of point objects moving on the sphere, and subsequently extend these to the scenario of circular patches moving on the surface of the sphere.  We use these conditions to determine (i) for fixed great circles of the objects motion, the speed ratios that will lead to collision, and (ii) for a given speed ratio and a fixed great circle of one of the objects, the set of great circles of the other object that will lead to collision.

\addtolength{\textheight}{-12cm}   




\end{document}